\documentclass[a4,10pt]{article}

\usepackage{makeidx}

\newif\ifpdf
\ifx\pdfoutput\undefined
   \pdffalse
\else
   \pdfoutput=1
   \pdftrue
\fi

\ifpdf
  \usepackage[
  pdftex,
  hyperindex,
  pagebackref,
  colorlinks,
  linkcolor=blue,
  menucolor=blue,
  pagecolor=blue,
  urlcolor=blue,
  citecolor=blue,
  pdftitle={PARFAIT: GNSS-R coastal   altimetry },
  pdfauthor={STARLAB BARCELONA SL},
  pdfcreator={LaTeX}
  ]{hyperref}

  \usepackage{thumbpdf}
  \usepackage{url}
  \usepackage[pdftex]{graphicx}
  \DeclareGraphicsExtensions{.pdf,.jpg,.mps,.png}

\else

  \usepackage[dvips]{graphicx}
  \DeclareGraphicsExtensions{.eps,.jpg,.mps,.png,.ps}

\fi

\usepackage{amsmath}
\usepackage{subfigure}  
\usepackage{fancyhdr}  
\usepackage{xfig}
\usepackage{multicol}

\graphicspath{{figures/}}

\usepackage[english]{babel} 

\usepackage[latin1]{inputenc}    % Aquest es el que s'encarrega de 
\catcode`\@=11

\catcode`\@=11
\def\section{\@startsection{section}{1}{\z@}{3.5ex plus 1ex minus
   .2ex}{2.3ex plus .2ex}{\large\bf}}

\def\eqnarray{\let\@currentlabel=\theequation\refstepcounter{equation}
    \global\@eqnswtrue
    \global\@eqcnt\z@\tabskip\@centering\let\\=\@eqncr
    $$\halign to \displaywidth\bgroup\@eqnsel\hskip\@centering
      $\displaystyle\tabskip\z@{##}$&\global\@eqcnt\@ne
       \hfil${{}##{}}$\hfil
      &\global\@eqcnt\tw@ $\displaystyle\tabskip\z@{##}$\hfil
       \tabskip\@centering&\llap{##}\tabskip\z@\cr}
\def\lefteqn#1{\hbox to 4\arraycolsep{$\displaystyle #1$\hss}}

\def\thesection{\arabic{section}.}

\def\appendix{\setcounter{section}{0}
        \def\thesection{Appendix.}
        \def\theequation{\Alph{section}.\arabic{equation}}}

\long\def\@makefntext#1{\parindent 0cm\noindent
\hbox to 1em{\hss$^{\@thefnmark}$}#1}

\newcommand\unmarkfootnote[1]{%
  \begingroup 
    \let\@makefntext\noindent
    \footnotetext{#1}%
  \endgroup}

\newcommand{\captionfonts}{\footnotesize}

\makeatletter  % Allow the use of @ in command names
\long\def\@makecaption#1#2{%
  \vskip\abovecaptionskip
  \sbox\@tempboxa{{\captionfonts #1: #2}}%
  \ifdim \wd\@tempboxa >\hsize
    {\captionfonts #1: #2\par}
  \else
    \hbox to\hsize{\hfil\box\@tempboxa\hfil}%
  \fi
  \vskip\belowcaptionskip}
\makeatother   % Cancel the effect of \makeatletter

\newsavebox{\fmbox}

\def\str{2003 Starlab Barcelona SL\space}

\begin{document}

\thispagestyle{empty}

\noindent
{\Large \bf PARFAIT: GNSS-R coastal altimetry}

\vspace*{0.3cm}
\noindent 
M. Caparrini, L. Ruffini, G. Ruffini\\
{\em Starlab, C. de l'Observatori Fabra s/n, 08035 Barcelona, Spain, http://starlab.es.} 

\unmarkfootnote{\copyright \str}

%%%%%%%%%%%%% Start abstract page %%%%%%%%%%%%%
\vspace*{0.2cm}
\section*{Abstract}
GNSS-R signals contain a coherent and an incoherent component. 
 A new algorithm for coherent phase altimetry over rough ocean surfaces, called  PARFAIT,  has been developed and implemented in Starlab's STARLIGHT\footnote{STARLab Interferometric Gnss Toolkit.} GNSS-R software package. In this paper we report our extraction and analysis of the coherent component of L1 GPS-R signals collected during the ESTEC Bridge 2 experimental campaign using this technique. The altimetric results have been compared with a GPS-buoy calibrated tide model with a resulting precision  of the order  1 cm.

%%%%%%%%%%%%% End abstract page %%%%%%%%%%%%%
\vspace*{0.2cm}

{\bf Keywords:} Passive radar, GNSS, GPS, Galileo, GNSS-R, GPS-R, altimetry, PARIS, PIP, PARFAIT, coastal applications.
\vspace*{0.5cm}

%\tableofcontents
\begin{multicols}{2}
%\section{The GPS reflected field retrieved by GPS processing}
%\label{ch:field_with_GPS}
\section{Introduction}
Specular reflections dominate medium to short wavelength electromagnetic forward scattering on the ocean, examples of which include GNSS and solar  reflections. 
As reported in \cite{ruffini2001a}, during the last decade many GPS-R (Global Positioning System Reflections) experimental campaigns have now been successfully carried out. A partial and surely incomplete list is provided in Table 1.  In this paper we focus on the potential of GNSS-R (Global Navigation Satellite System Reflections) for altimetric coastal applications. The techniques developed, however, can also be implemented in other scenarios (airborne, spaceborne).

 The specularly scattered field is  composed of a coherent component and a random, Hoyt-distributed component \cite{beckmann1963}. When the surface is very rough, the latter becomes incoherent and the former becomes very small.  In fact, if the surface height distribution is normal with deviation $\sigma_\zeta$,  then 
\begin{equation}
\langle r^2 \rangle \sim n^2 e^{-( 4\pi \sigma_\zeta\cos\theta /\lambda )^2} +
\end{equation}
$$
 \; \; \; \; \;\; \; \; \; \;\; \; \; \; \; \; \; \; \; \; \; \;  n(1- e^{-(4\pi \sigma_\zeta\cos\theta /\lambda)^2} ),
$$
where $\langle r^2 \rangle$ is the power average, $n$ is the number of scatterers, $\lambda$ the EM wavelenght and $\theta$ the local incidence angle
%Thus, we see that the magnitude of the reflected field should depend mainly on the ratio of significant wave height to wavelength 
\cite{ruffini2001a,ruffini1999c}.  

GNSS-R signals thus contain a coherent and an incoherent component. In companion papers we discuss the analysis of the incoherent component for sea state monitoring \cite{germain2003,ruffini2003b} and for code altimetry \cite{ruffini2003}. Here we present a new approach for the extraction and analysis of the coherent component of GNSS reflected signals to perform phase altimetry.

The data discussed here was collected by ESA/ESTEC during the  Bridge-2 experiment. The experiment aimed at gathering direct and reflected GPS signals from antennas located about 18 m above the mean sea level of an Estuary in the North sea of Holland. For more information on the experimental setup, the reader is directed to \cite{belmonte2003}.

This paper is structured as follows:
\begin{itemize}
\item Discussion on coherence properties of reflected signals and their use for  phase altimetry.
\item Analysis of the direct and reflected signals and EM field extraction.
\item Implementation of PARFAIT altimetry.
\item Comparison with other data and discussion of the altimetric results.
\end{itemize}

\begin{table*}
\hspace{-1cm}
 \label{tab:GNSSR_experiments}
\centering{

   \begin{tabular}[tb]{|p{1.6cm}c|p{1cm}|c|p{7cm}|}
     \hline
 Main \hspace{5mm} Author & & Inst. & Date & Notes \\
        \hline \hline
  Garrison &\cite{garrison1996} & JPL   & 1996  & A normal GPS receiver was used. Demonstrated reception/tracking  of reflected signals over relatively calm waters.  Concluded  that more complex receiver is needed.  \\
       \hline          
   ${\mbox{Martin-Neira}}$ Caparrini&\cite{martin-neira2001,caparrini1998}& ESA & 1997  & GNSS-R PARIS altimetric experiment from a Zeeland bridge.  C/A code  used for correlation leading to an altimetric accuracy in the order of 3 meters after 1 second (1\% of the chip length).   \\
       \hline          
  Garrison & \cite{garrison1998}& JPL & 1997 &  Widening of correlation function in rough seas demonstrated. Application for sea state from air. \\
       \hline          
  Komjathy &  \cite{komjathy1998}& CCAR & 1998 &  Aircraft experiments, 3-5 km altitude. \\
       \hline          
 LaBrecque & \cite{labrecque1998}&NASA & 1998 &  The first spaceborne observation of GPS signals reflected from the ocean surface.\\
       \hline          
 Cardellach Ruffini Garrison& \cite{cardellach2003,garrison2000}& IEEC & 1999 & Balloon experiment. Successful detection of reflected signals at 38 km of height with low gain antenna. Sea surface winds retrieved with $\sim$2 m/s error.   \\
        \hline
 Cardellach Ruffini& \cite{cardellach2001a}& IEEC & 1999 & First ESA aircraft experiment. Some signals detected, DDM produced,  but experiment failed due to hardware problems.  \\
       \hline 
 Armatys & \cite{armatys2000} & CCAR & 2000 & Wind speed and directions obtained from reflected GPS signals are compared to the SeaWinds scatterometer on-board QuikSCAT.\\
     \hline
 Garrison & \cite{garrison2000a} & JPL & 2000 & With GPS-R airborne data, retrieval of the wind speed with a bias of less than 0.1 m/s and with a standard deviation of 1.3 m/s.\\
     \hline
Zavorotny & \cite{zavorotny2000}&  CCAR & 2000 & Fundamental theoretical work. Comparison of experimental and theoretical waveforms.\\
     \hline
Zuffada & \cite{zuffada2000}& JPL & 2000 & Lakeside experiment, with an almost flat surface (no roughness). Centimetric phase altimetry.\\
     \hline
${\mbox{Martin-Neira}}$, Ruffini, Serra, Colmenarejo & \cite{ruffinipipaer2000}& ESA& 2000 & The pond experiment was designed to test some key issues in the PARIS Interferometric Processor (PIP) concept. The PIP concept is based on the use of dual-frequency carrier measurements to exploit the correlations in the scattered signals at similar frequencies.\\
     \hline
 Ruffini \hspace{5mm}Caparrini & \cite{caparrini2001} & Star-lab IEEC & 2001 & GPS-R L1 data collected from the Casablanca drilling platform by IEEC has been analysed at Starlab.\\
     \hline
${\mbox{Martin-Neira}}$ & \cite{belmonte2002}& ESA & 2001 & The experimental campaign which is the object of this work.\\
     \hline
Cardellach Starlab Team&\cite{cardellach2002,cardellach2002b,cardellach2002c,germain2002}& IEEC/ Starlab& 2001 & GPS-R data collection from airborne platform. Campaign performed within the  ESA/ESTEC project ``PARIS Alpha''. Data processed under ESA/ESTEC projects ``PARIS Alpha'' and ``OPPSCAT 2''.\\
      \hline
Starlab Team &\cite{germain2003,ruffini2003} & Starlab & 2002 & GPS-R data collection from airborne platform. Retrieval of altimetric profile matching Jason-1. (ESA/ESTEC ``PARIS Gamma'')\\
  \hline
Starlab Team &\cite{ruffini2003b} & Starlab & 2003 & GPS-R data collection from Barcelona Harbour (HOPE campaign, Starlab Oceanpal project).\\
      \hline
%\caption{A list of the most significant GPSR experimental campaigns.}
\end{tabular}}
\vspace{.5cm}
\caption{Representative GNSS-R experiments and milestones (1996-2003).}
\end{table*}

\section{The  complex field}
The importance of retrieving the coherent part of the EM field backscattered by the sea surface  stems from  its altimetric content. Measuring the phase of the coherent component allows for accurately estimating the delay of the reflected signal with respect to the  the direct one, i.e., for  estimation of the temporal {\em lapse}. This is the essential measurement for  altimetry  \cite{ruffinipipaer2000}.

In order to collect the complex EM field, the complex signal is generated from the real one. Although this operation is often performed by the receiver front-end, in the Bridge-2  experiment only the in-phase component of the signal was sampled at high frequency and stored on digital tape. The quadrature component was generated afterwards. The process is illustrated in Figure~(\ref{fig:downconvertion}).
We can then represent the direct signal received at the antenna input\footnote{This contains only the C/A code part. The P code component can be neglected thanks to the subsequent correlation of the signal with replicas of the C/A code---the two codes are  orthogonal.} as
\begin{equation}
  S_d(t)=A_d\cdot C(t)\cdot D(t)\cdot e^{i(\omega_{L1}+\omega_d) t} + \eta_d,
\end{equation}
where $A_d$ is the direct signal amplitude, $C(t)$ represents the C/A code, $D(t)$ the navigation code, $\omega_{L1}$ the L1 carrier frequency,  $\omega_d$ the Doppler frequency offset, and $\eta_d$ (thermal) noise. The reflected signal at low altitudes can be modelled by
\begin{equation}
  S_r(t)=C(t)\cdot D(t)\cdot e^{i(\omega_{L1}+\omega_d )t}\cdot
\end{equation}
$$
 \; \;\; \; \; \; \;\; \; \;\; \; \; \left(A_r\cdot e^{2\pi i{{\cal L}/\lambda}}+O(t)\right)+\eta_r,
$$
where $A_r$ is the reflected signal mean amplitude, $O(t)$ represents the perturbation due to ocean motion and $\cal L$ the reflected signal extra path length. In coastal applications $O(t)$   is a relatively slowly varying  quantity with zero mean, while $\cal L$, which contains the geophysical tide signal, can be considered effectively frozen during correlation processing.

 After modulation with a local oscillator of frequency $\omega_{L1}-\omega_{IF}$ and low-pass filtering, the signal will have a residual carrier at $\omega_d+\omega_{IF}$. This signal is  mixed with a phasor at frequency $\omega_{IF}+\tilde{\omega}_d$, where $\tilde{\omega}_d$ is an estimate of the Doppler frequency for the satellite under investigation, and finally low-pass filtered. 
%The result of such a mixing is a signal with two frequency component: $\Delta\omega_d$, the error in the Doppler frequency estimation, and $2(\omega_{IF}+\omega_d)-\Delta\omega_d$, a spurious frequency which is not possible to filter out and that is considered as a noise component.

With the assumption that the navigation bit is constant during the integration time (which is correct if the correlation is bit-aligned and the coherent integration time $T_{E}$ is less than 20 ms), and considering that during an integration time interval the value of $\Delta\omega_{d}$ is constant,  the complex  \em p-th \rm sample of the correlation coefficient for the direct signal writes
\begin{eqnarray}
C_{p}\,&\sim &\,\frac{1}{2} A_d\, D_k  \,R_p\, e^{-i\Delta\omega_{d_p}T_E\left(p-1\right)}\cdot  \nonumber \\ 
&\cdot&e^{-i\Delta\omega_{d_p}\frac{T_s}{2}}\frac{\sin\left(\frac{\Delta\omega_{d_p}}{2}T_E\right)}{\sin\left(\frac{\Delta\omega_{d_p}}{2}T_s\right)}, 
\label{eq:exp_corr2}
\end{eqnarray}
where  $T_s$ is the sampling interval and $R_p$ the corresponding correlation coefficient function. For the reflected signal we can write an equivalent expression, modulated by the slowly varying phasor $A_r\cdot \exp(2\pi i{\cal L}/\lambda)+O(t)$. For coastal applications we can assume there will be little filtering of this quantity by the coherent integration process, as the ocean moves slowly compared to coherent integration times (a few ms).

In the case of the direct signal, we can  easily track  the carrier phase. To this end, the delta-phases obtainable from equation \eqref{eq:exp_corr2} can be accumulated using 
\begin{eqnarray}
\phi_{p+1}&-&\phi_p = \nonumber \\
&=&\mbox{Im}\left(\log{\frac{C_{p+1}}{C_p}}\right)\,=\,-\Delta\omega_{d_p} T_E.
\label{eq:delta_phi}
\end{eqnarray}
This equation holds while $\Delta\omega_{d_{p+1}}\approx\Delta\omega_{d_p}$. This is a good approximation, since the time during which this variation is measured is the  coherent integration time.

The main advantage of using this algorithm  for phase tracking is that, due to its differencing nature, it allows  for easy detection of the navigation bit  $\pi$ radians phase change.

\begin{figure}[tbhp]
\centering
        \includegraphics[width=6cm]{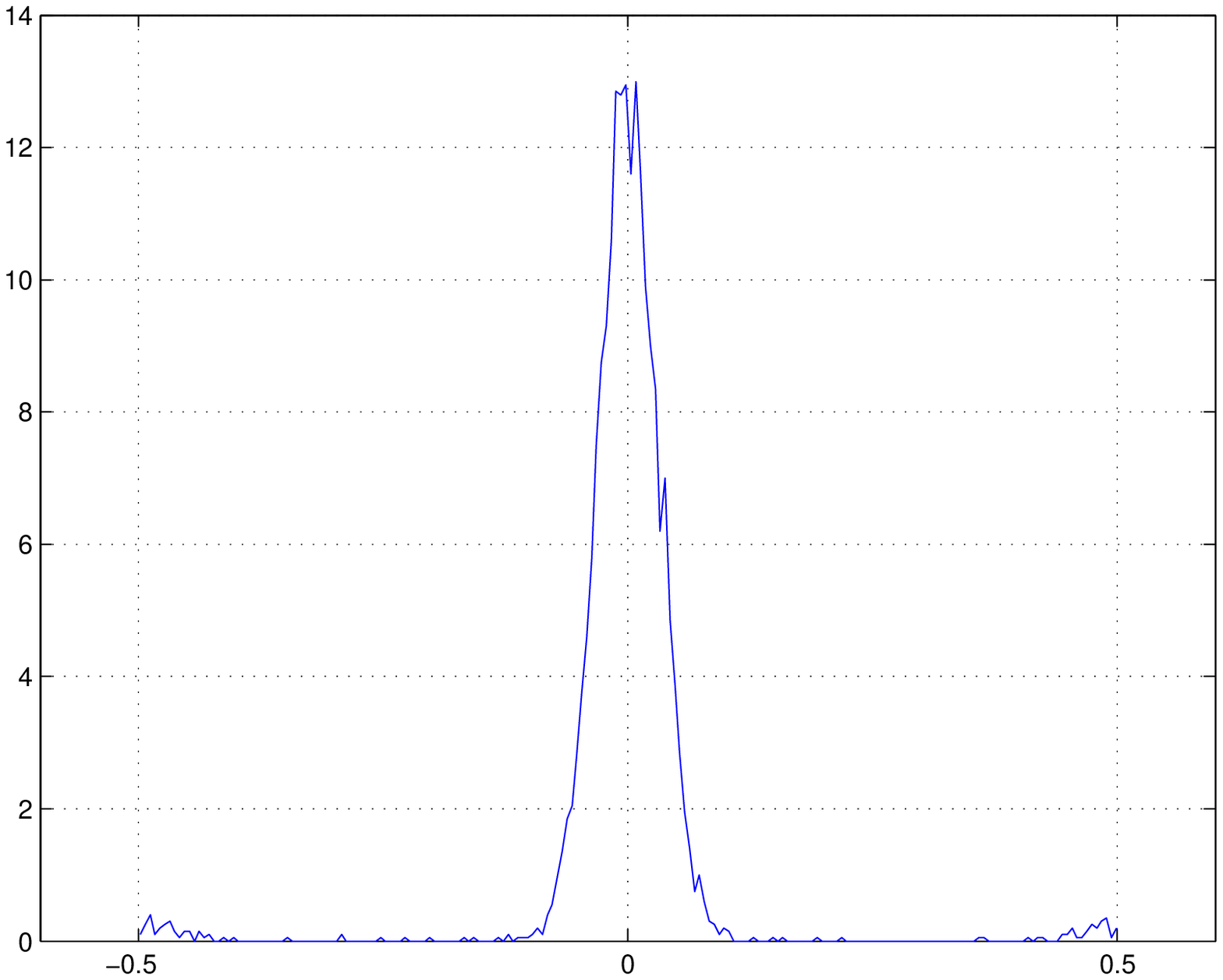} 
    \caption{The histogram of the carrier phase variation, measured on an integration time interval (in this case 1 ms). The x-axis represents cycles, while on the y-axis there are arbitrary units. The accumulation of $\delta$-phase values around zero can be seen, as well as in the vicinity of $\pm$ half a cycle.}
  \label{fig:example_deltaphase_hist}
\end{figure}

\begin{figure}[tbhp]
\centering
        \includegraphics[width=6cm]{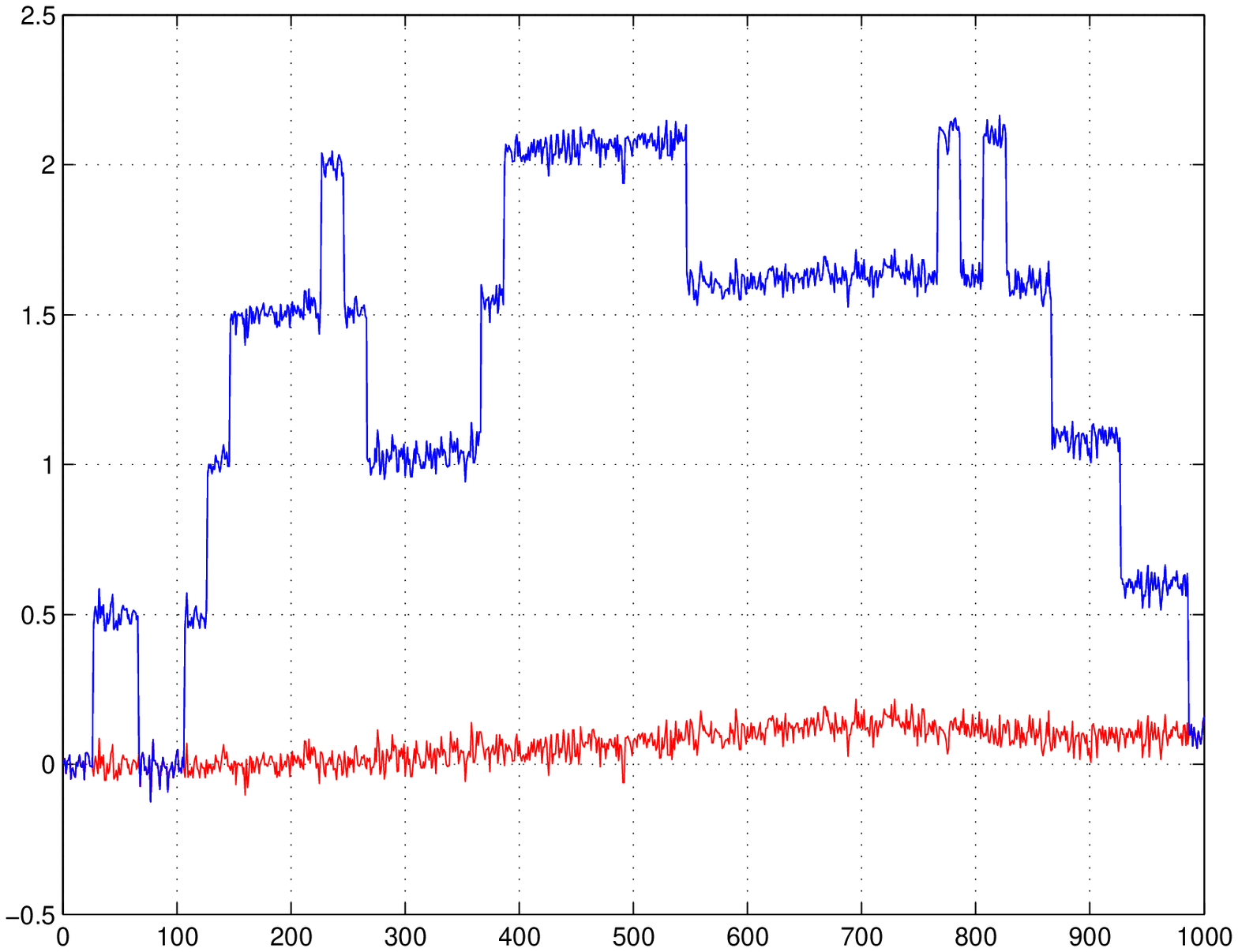} 
    \caption{The directy signal carrier phase obtained accumulating the $\delta$-phase according to Equation \eqref{eq:delta_phi}. The stepped plot represents the accumulated phase \em as it is \em, i.e. without compensating for the navigation bit half-cycle variation, which is clearly visible. The lower curve represents the same phase after removal of this effect (navigation bit correction). The units on the x-axis are milliseconds, and on  the y-axis thgy are  cycles.}
  \label{fig:example_phase_uphase}
\end{figure}

\begin{figure}
  \centering
    \subfigure[Direct field.]
    {\includegraphics[width=6cm]{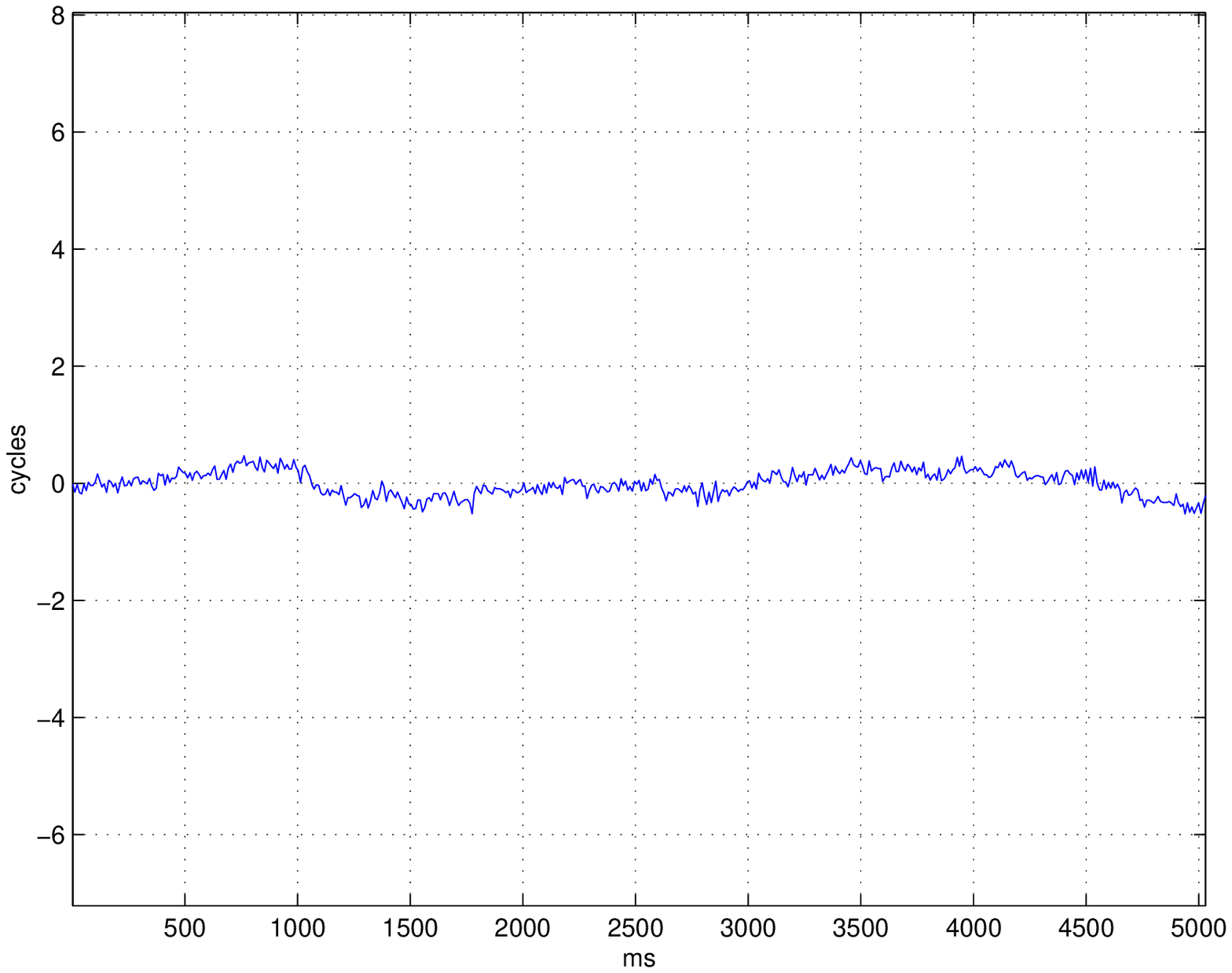}
      \label{fig:casab_phase_detrend_dir}}
    %\hspace{1cm}
    \subfigure[Reflected field.]
    {\includegraphics[width=6cm]{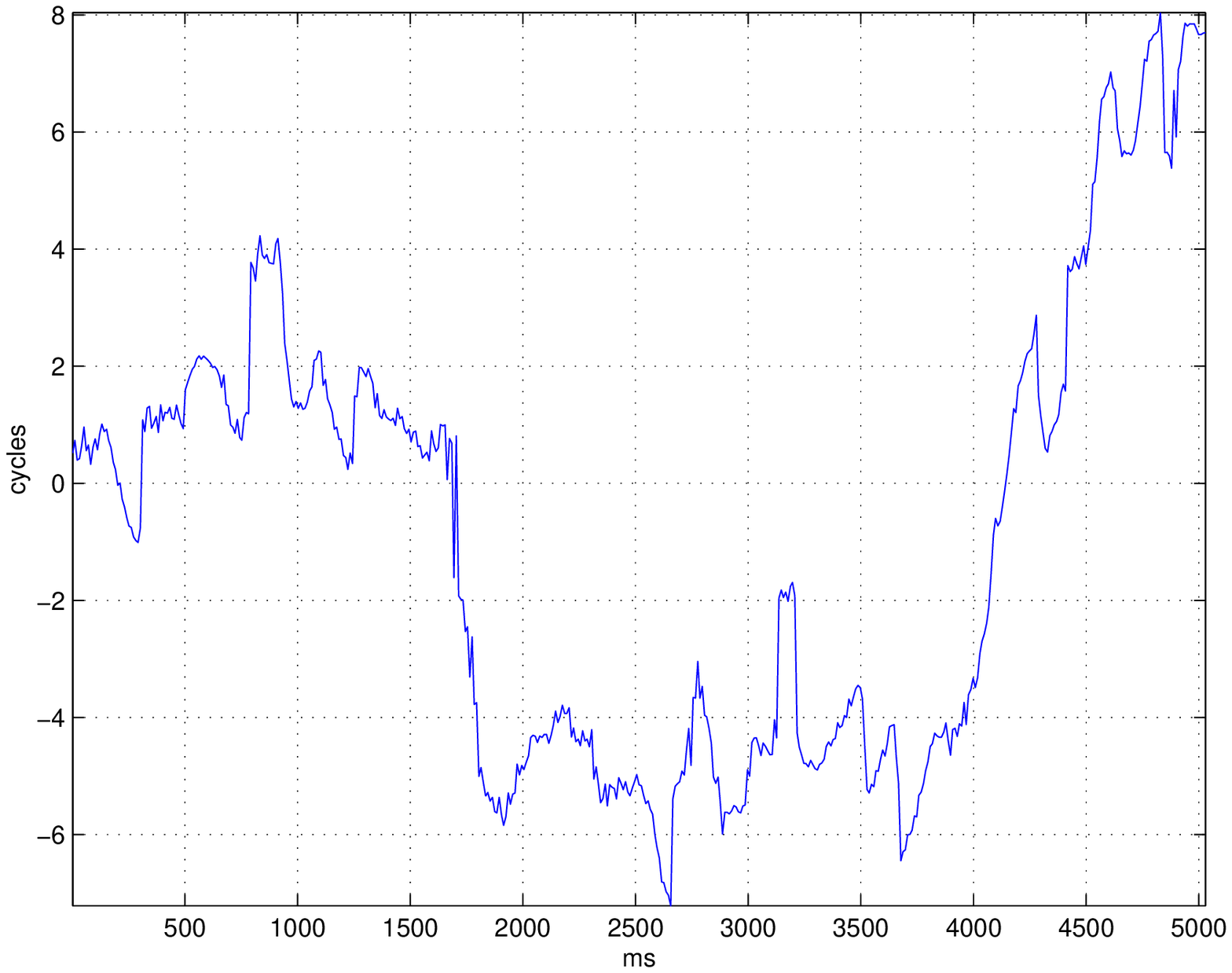}
      \label{fig:casab_phase_detrend_ref}}
  \caption{Example of the tracked phase, without the Doppler contribution. The units are milliseconds on the x-axis and cycles on the y axis. The integration time is 20 ms.}
  \label{fig:casab_phase_detrend}
\end{figure}

\begin{figure}
  \centering
    \subfigure[Direct field.]
    {\includegraphics[width=6cm]{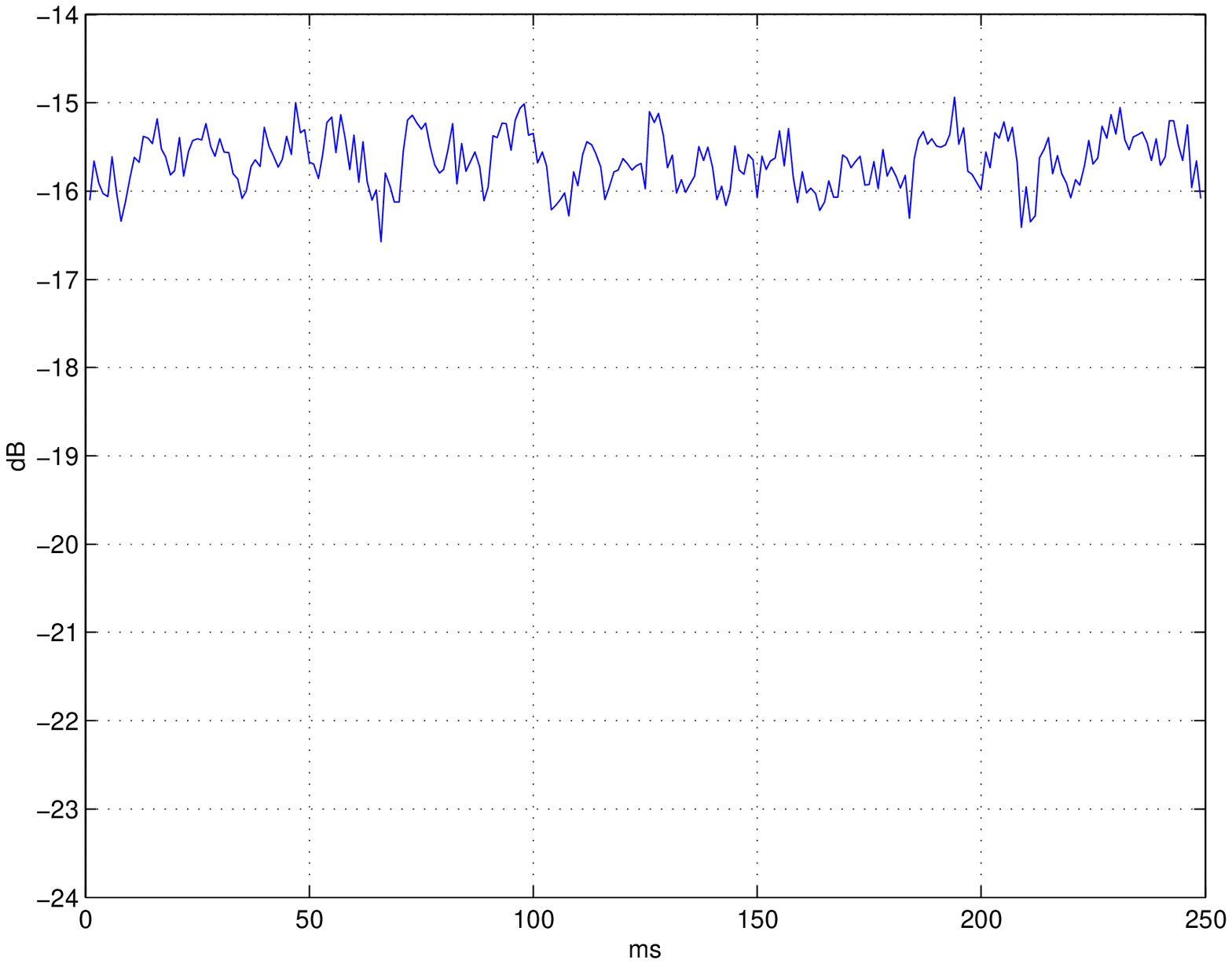}
      \label{fig:casab_ampl_dir}}
    %\hspace{1cm}
    \subfigure[Reflected field.]
    {\includegraphics[width=6cm]{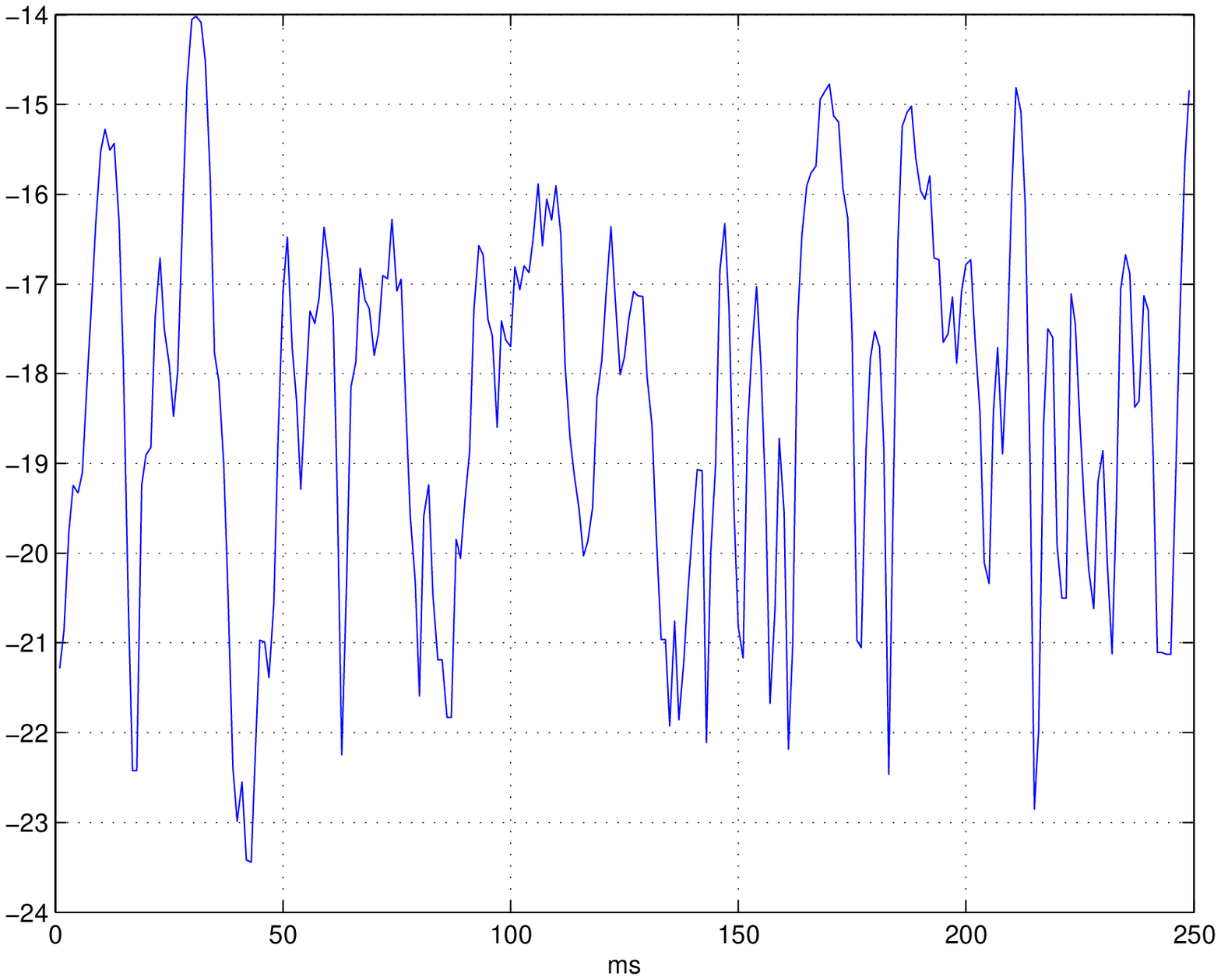}
      \label{fig:casab_ampl_ref}}
  \caption{Example of field amplitude time series (phasor dustball, Casablanca Experiment). The units are milliseconds on the x-axis and correlation coefficient units in dB on the y axis. The integration time is 20 ms.}
  \label{fig:casab_ampl}
\end{figure}

\begin{figure}
  \centering
    \subfigure[Direct field.]
    {\includegraphics[width=6cm]{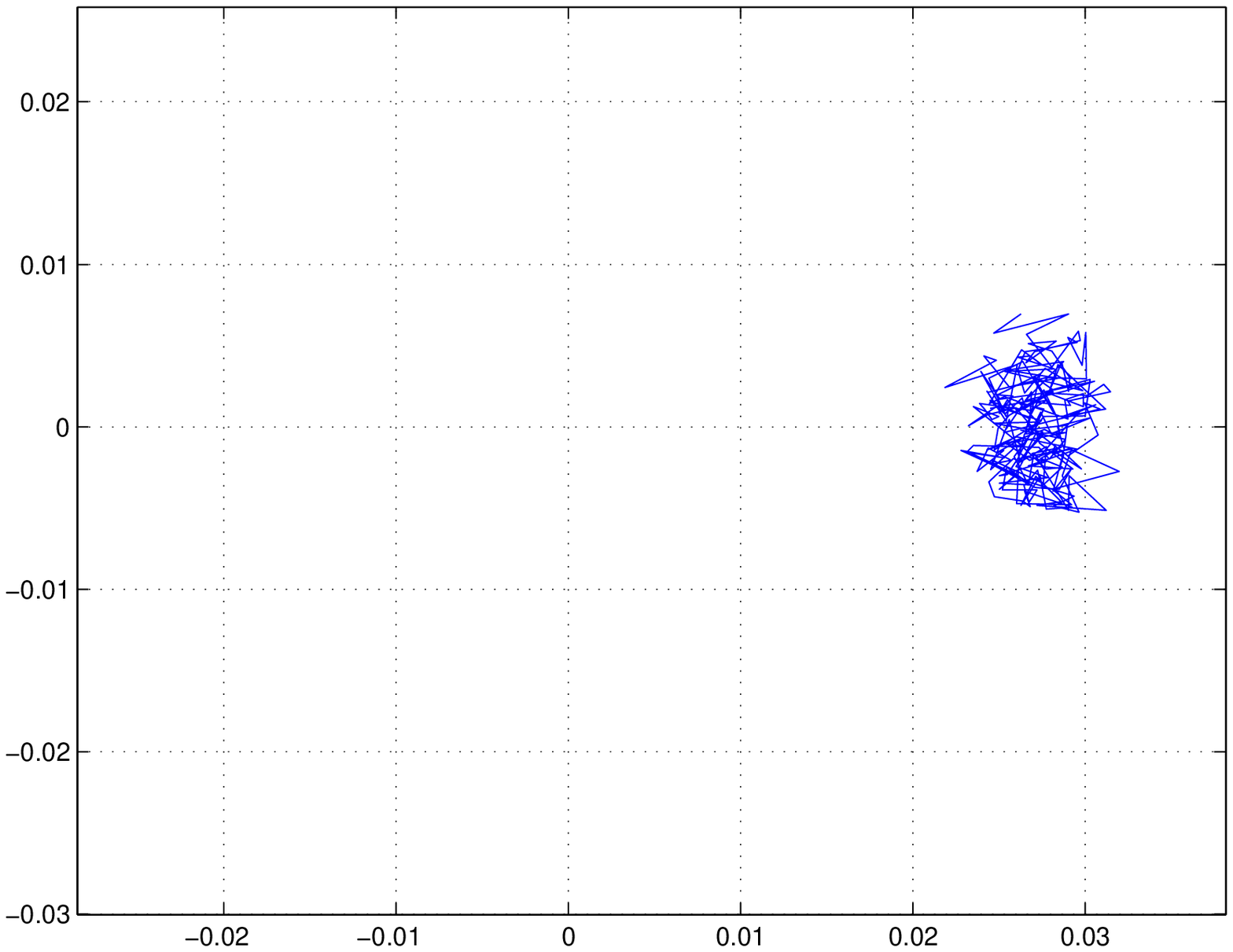}
      \label{fig:casab_ampl_detrend_dir}}
    %\hspace{1cm}
    \subfigure[Reflected field.]
    {\includegraphics[width=6cm]{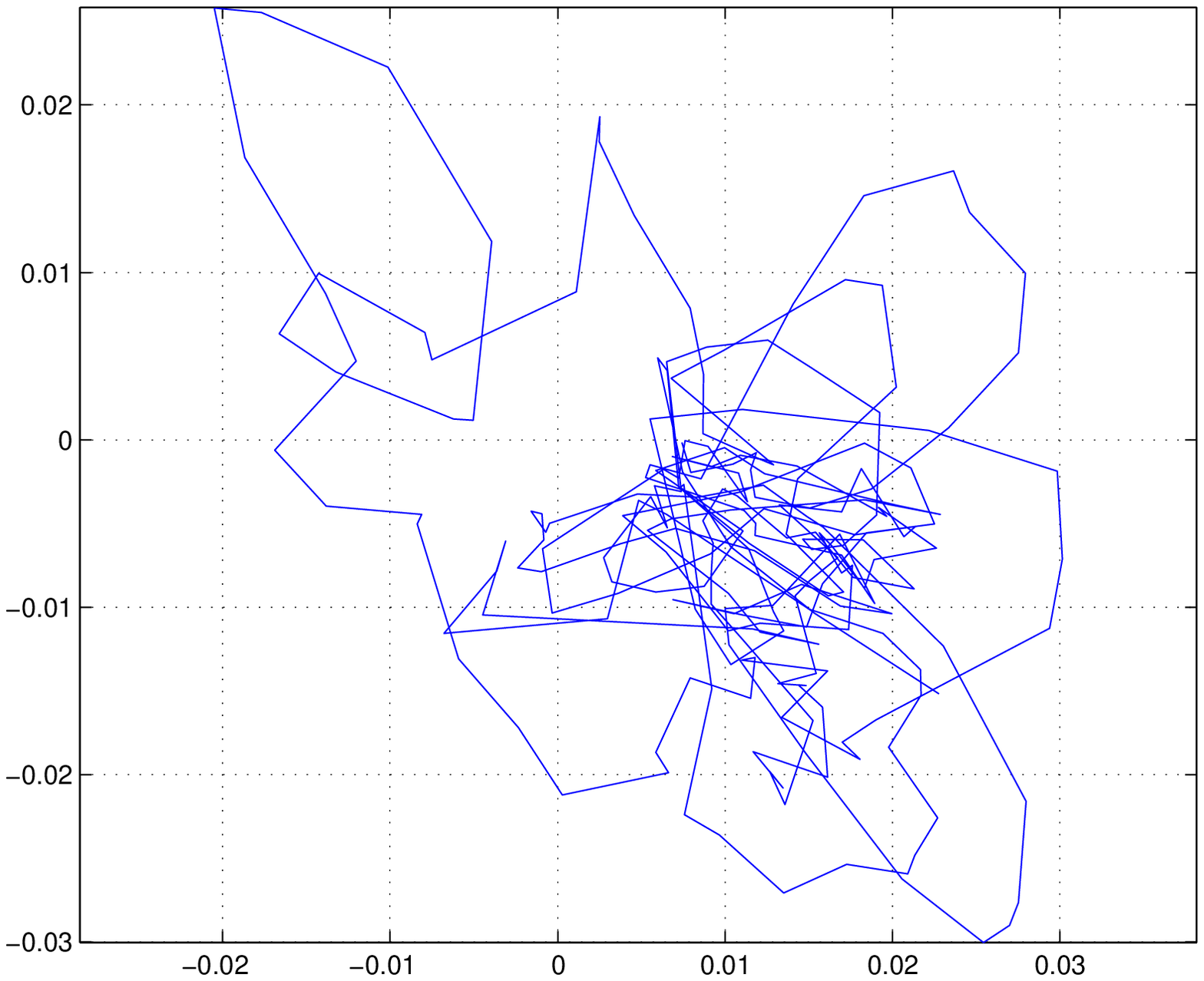}
      \label{fig:casab_ampl_detrend_ref}}
  \caption{Example of complex field time series (phasor dusball, Casablanca Experiment). The units are correlation coefficient units on both axis. The coherent integration time is 20 ms.}
  \label{fig:casab_ampl_detrend}
\end{figure}

\begin{figure}[tbhp]
\centering
        \includegraphics[width=10cm]{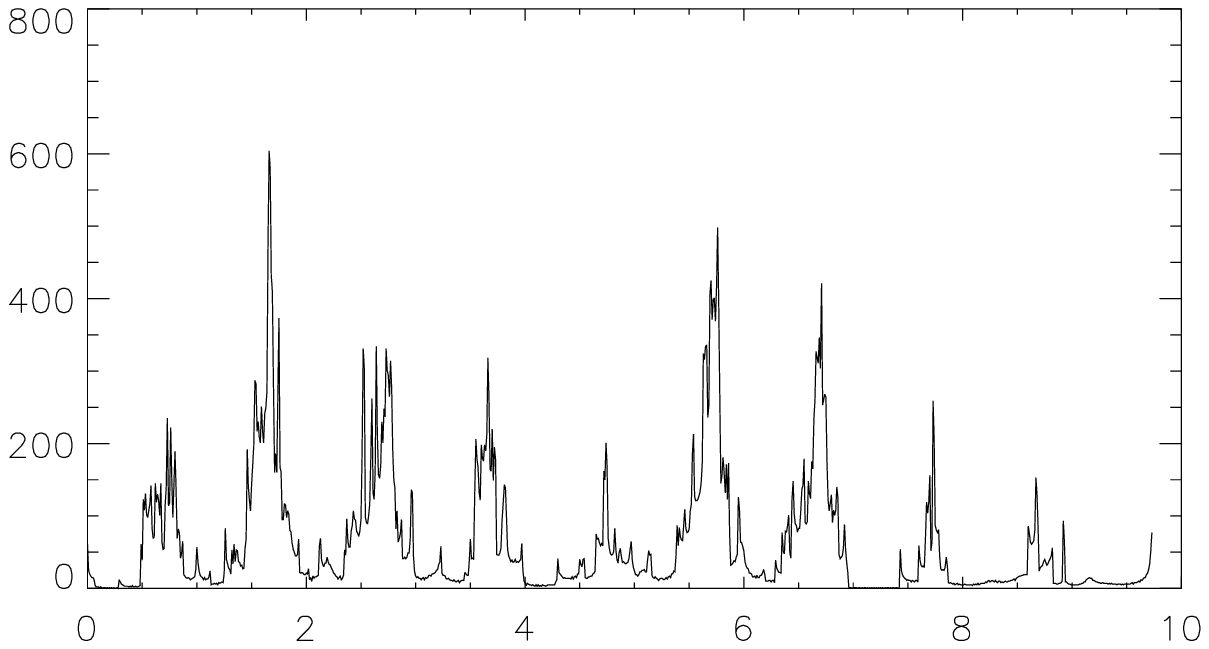} 
    \caption{Histogram of the phase for a noiseless simulated field. It can be observed that the unwrapped phase wanders around multiple winding number kingdoms, while tending to spend more time around an average complex field point. This illustrates the fact that the unwrapped phase cannot be used directly for altimetry, even in the absense of noise.}
  \label{fig:phasehistogram}
\end{figure}

\begin{figure}[tbhp] 
\centering
        \includegraphics[width=10cm]{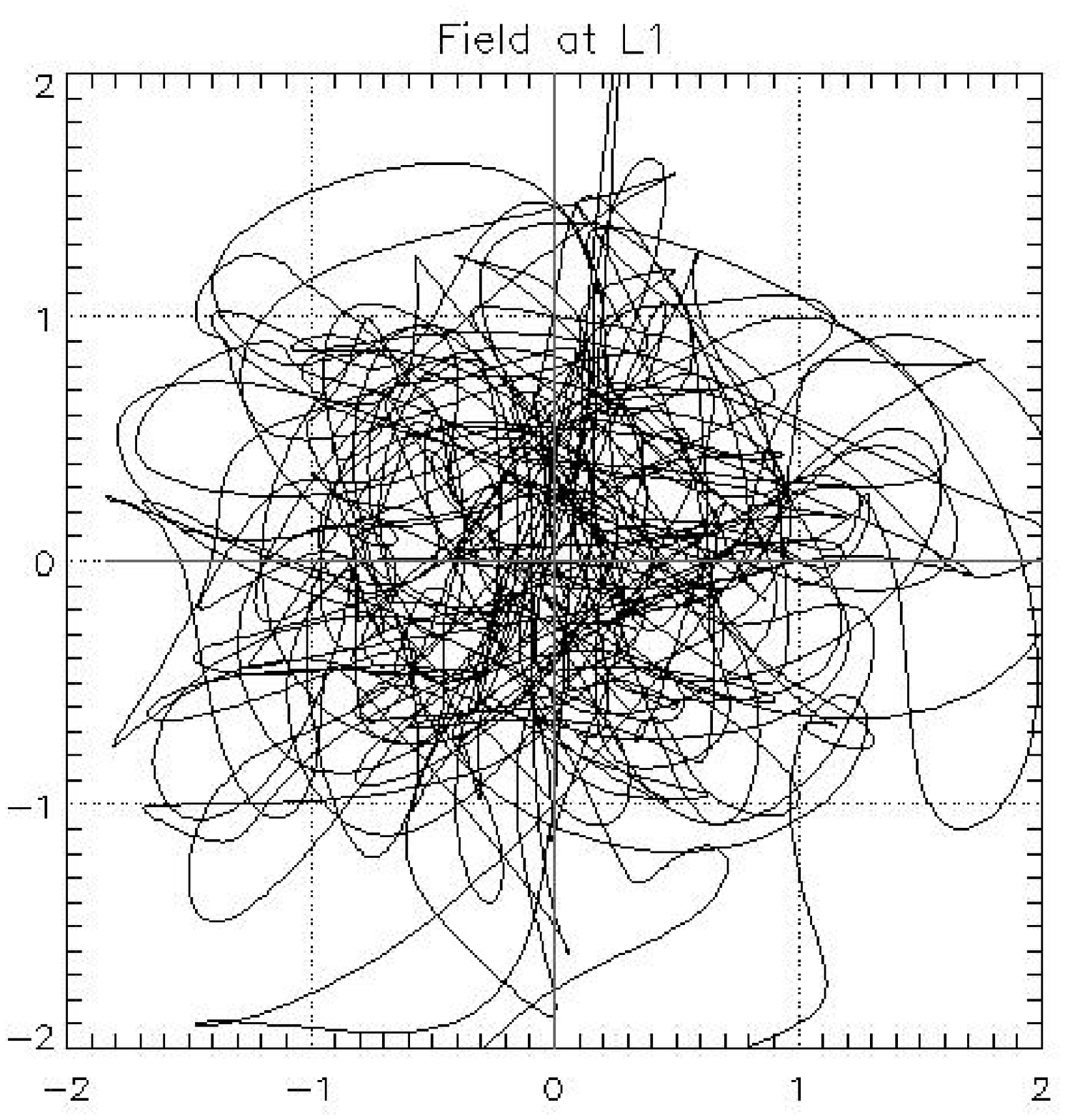} 
    \caption{The simulated EM field at L1 frequency (phasor dustball), after reflection on the sea surface.  The simulation has been performed with the GRADAS software \cite{ruffini2001b} developed by Starlab. This simulation is for a wind speed of $U_{10}=3$ m/s.}
  \label{fig:pacos_dustball}
\end{figure}

Figures  \ref{fig:example_deltaphase_hist} to \ref{fig:pacos_dustball}, which illustrate these concepts, refer to the processing of another set of GPS-R data---collected during the Casablanca oil platform Experiment. This Repsol owned drilling platform is about 40 km off the coast of Tarragona, Spain ($40^o 43' 4'' N$, $1^o 21' 34'' E$). The measurement campaign took place on March 14th, 2000.

In Figure \ref{fig:example_deltaphase_hist}, the histogram of the delta-phases is shown. The x-axis represents cycles and the y-axis is in  arbitrary units. Most $\delta$-phase values are clearly concentrated around zero. Other values gather just before $\pm\pi$. These values represent in fact \it small \rm values to which $\pm\pi$ radians have been added on occurrence of a navigation bit transition. 

In Figure \ref{fig:example_phase_uphase}, the direct signal phase with and without  navigation bit correction is plotted.
 In Figure \ref{fig:casab_phase_detrend}, the phase for the (navigation bit corrected) direct and reflected signals is shown. The effect of the reflection on the sea surface is clearly visible in the large variations present in Figure~\ref{fig:casab_phase_detrend_ref} with respect to Figure~\ref{fig:casab_phase_detrend_dir}.
In Figure~\ref{fig:casab_ampl} and \ref{fig:casab_ampl_detrend}, the amplitude magnitude and  the complex vector of the direct and reflected fields, respectively, are shown. In Figure~\ref{fig:pacos_dustball}, a simulation of the L1  GPS complex field phasor {\em dustball} after reflection, akin to the one in  Figure~\ref{fig:casab_ampl_detrend_ref}, is shown. The simulation parameters  have been  chosen to match the Casablanca experiment sea state. Those were reported as a ``quite calm sea with a gentle breeze'', with SWH of about 0.7 m as measured by a nearby buoy.

\section{The PARFAIT approach}
\label{ch:diff_approach}
%\subsection{Introduction}
In general, the  altimetric information content in the PARIS interferometric field phase will be very difficult to use. This is due to the impact of the incoherent component in the reflected signals. The incoherent component causes fading and winding. 

On the one hand, at a practical level, fading events will prevent stable phase tracking of the complex field. Even if as in  in the Bridge-2 experiment  the sea surface is relatively smooth and fading events are not so frequent,  a single event can severly complicate the use of phase information if countermeasures are not taken. In general, however,  the reflected field will fade  very often. As discussed in  \cite{belmonte2003}, it is possible to inject in the system (during a fading event) a model-based phase to ``glue'' the phase history, but this approach will in general necessitate the input of too much model information into the data in rough sea conditions.

More importantly, as explained in the previous section, the reflected field  incoherent component will cause arbitrary winding of the field phasor. This means that the reflected unwrapped phase, unlike the direct one,  cannot be directly used for ranging. Indeed, as we have shown in previous work \cite{ruffinipipaer2000}, the reflected field accumulated phase will generally  wander around the complex plane, travelling  to different winding number kingdoms, {\em even in the absence of fadings} (see Figure~\ref{fig:phasehistogram}). That is to say, even if a very high SNR system is devised to get around the problem of field fadings, the interferometric unwrapped phase will not be directly usable for ranging. Unlike the problem of fadings, this is a fundamental issue, not a practical one. 

An approach discussed in \cite{ruffinipipaer2000,martin-neira2003}, PIP\footnote{PARIS Interferometric Processor.},
%\footnote{PARIS Interferometric Processor, US patent....XXXXXXXXXXXX}, 
involves the use of multiple frequencies for the synthesis of a longer wavelength which will be more immune to fadings. Here we discuss another approach, PARFAIT\footnote{PARFAIT stands for PARis Filtered-field AltImetric Tracking.}, which is in fact complementary to PIP. 

%Another possibility is not to track the reflected field phase at all but to track the direct field minus the reflected field phase, which is really what PARIS concept is about. \\

%et us summarize why we should not try to track the reflected field phase at all.
%Using the  phase for altimetry presupposes first the possibility of  track the reflected field phase (which is difficult to the event of fadings), and second, that the reflected field phase contains useful ranging information (which is in general false, due to dominance of the incoherent field part). We note that the second point is often overlooked. Yet we know that the accumulated phase acts like a random walk. The average accumulated phase is very badly behaved (essentially meaningless). \\

%In PARFAIT, we first note that tracking the reflected phase is important if the signal of interest rotates fast  compared to noise-induced rotation. This is definitely not the case in the bridge phase or in any other coastal application of the technique. In fact, it is never true for PARIS applications over the oceans. It is also very difficult to do: errors accumulate fast, as above mentioned for example, during fading events.
In the PARFAIT approach, we begin by noting that although  the  reflected field unwrapped phase carries no ranging information, this need not be  a  fundamental problem. What is needed  is the coherent {\em geophysical} field component in the signals, which is near zero frequency in comparison with the others---a sort of  average field. 
%This exists even if the phase wanders and wanders. 
This average field is just the coherent component in the reflected signals after downconversion. With this in mind, PARFAIT consists of the three steps described next.

The first practical step to extract the coherent part is to work with the interferometric field, the ratio for reflected versus direct complex field. This has the advantadge of error cancellation, e.g., in Doppler matching of the incoming signals, and of depending only on the lapse.

The second step is to ``counter-rotate'' the interferometric field using an a-priori model of the reflection process.

The third step is filtering the resulting counter-rotated interometric complex field to finally extract the coherent phase for estimation of the carrier lapse phase. Counter-rotation allows for longer filtering times. These are fundamental to extract the coherent component, which decays exponentially with the square of  sea surface standard deviation (sea state) over effective wavelenght (the wavelength divided by the sine of satellite  elevation). 

Finally, the phase lapse information obtained from the couter-rotated, averaged, complex interfereometric field is used for altimetry.

This  new approach to PARIS altimetry is described in more detail in the following sections.  As we discuss, it has proven to be a very robust and precise processing method.\\
\section{PARFAIT processing of the Bridge 2 dataset}
\label{parfait_proc}
%In this section the basis of the \sc parfait \rm  processing are presented. The \sc parfait \rm  processing is an innovative processing for GNSS-R data developed by Starlab, as part of the STARLIGHT GNSS-R software package. In the frame of this CCN, the application of such a processing will lead to the estimation of the height of the antennas of the Bridge 2 Experiment over the sea surface.\\

At low altitudes, simple geometrical considerations lead to the following equation relating the height of the receiver over the reflecting surface (considering the same height for the upward looking antenna and for the downward looking antenna) with the lapse---the measured delay measured between the direct and reflected GNSS signals:
\begin{equation}
{\cal L}_P(t)= c\Delta\tau_{P}\left(t\right) =2h(t)\cdot \sin\left(\epsilon_{P}\left(t\right)\right) +{b},
\label{eq:delayVSh}
\end{equation}
where ${\cal L}_P(t)$ is the lapse in meters at time $t$, $c$ is the speed of light, $\Delta\tau_{P}(t)$ is the temporal lapse in seconds, $h(t)$ is the height of the bridge, $\epsilon_{P}(t)$ is the elevation of the GPS satellite with a specific PRN number $P$, and $b$ is the hardware-induced delay bias, considered to be a common constant in time.
A first estimation of the height of the receiver can easily be performed through a linear fit of the lapse with respect to the sine of the elevation angle of each satellite.

In phase processing, the lapse is measured only up to an integer number of cycles $N$. Equation \eqref{eq:delayVSh} must be rewritten as follows
\begin{equation}
{\cal L}^c_P(t)=2h(t)\cdot \sin\left(\epsilon_{p}\left(t\right)\right) +{b} +N_p \lambda,
%delay_{p}\left(t\right)\,&=&\,2\,h\left(t\right)\,\sin\left(\epsilon_{p}\left(t\right)\right)+\nonumber \\
%& &+N_{p}\lambda+{b}
\label{eq:delayVSh_phase}
\end{equation}
with ${\cal L}_{p}^c\left(t\right)$ is the carrier lapse in meters and $\lambda$ is the carrier (L1) wavelength. In other words, the equation for  each satellite contains an additional unknown parameter, $N_{P}$. In order to use all the satellites for one height estimation, it becomes necessary to estimate also $N_{P}$, i.e. to solve the ambiguity problem. 

%The following procedure was used. All the measured ${\cal L}_{p}(t)$ have been put in an unique vector, with an additional additive parameter $b_{p}$
%\[ ({\cal L}_{1}(t_1),\,{\cal L}_{1}(t_2),\,...,\:({\cal L}_{2}(t_1)+b_2), ... \]
%\[ \, ({\cal L}_{2}(t_2)+b_2), \:...,\:({\cal L}_{p}(t_m)+b_p)).\]

In order to solve the estimation problem, a minimization search is carried out for all these paremeters: $h$ and $b$ (as real constants) and $N_P$ (as integers).

% The corresponding vector for the sine of the elevation angles was then built. Using these two vectors, a linear fit has been performed, resulting in a certain value for the norm of the residuals of the fit. Finally, considering this norm as a function of the parameters $b_{p}$, the optimum value $\hat{b}_{p}$ that minimises the norm has been found. For this  $b_{p}$ one obtains the best interpolation, i.e. the best value for a first estimation of the height of the bridge.\\
%This estimation of the height is nonetheless quite rough. Let us see why and how to improve this estimation.
%First of all, remember that the term on left-hand side of equation \eqref{eq:delayVSh_phase} can be written as (see also equation \eqref{eq:interf_field})
%\begin{equation}
%{\cal L}_{p}\left(t\right)\,=\,\lambda d\phi_i^{R-D}
%\end{equation}
%where $d\phi_i^{R-D}$ is the phase of the (reflected minus direct) interferometric field. This field is generally corrupted by fadings and, during fadings, the phase of this interpherometric signal is completely impossible to track (basically, during these events, the reflected signal is not present at all). Moreover, the optimisation previously described is performed in the $\mathcal{R}$ domain, without correctly approaching the ambiguity problem which is inherently to be solved in the integer domain.

However, as discussed,  the interferometric field should be first filtered to extract its coherent component. Filtering should be long enough to extract the coherent component but short enough to keep the geophysical signals of interest pass through.% Consider the usual equation
%\begin{eqnarray}
%\lambda d\phi_i^{R-D}\left(t\right)\,&=&\,2\,h\left(t\right)\,\sin\left(\epsilon_{p}\left(t\right)\right)+\nonumber \\
%&+&{b}.
%\end{eqnarray}
%The value of $d\phi_i^{R-D}$ 

This means that the geophysical coherent component we are after should not change for more than a small fraction of a cycle during the time duration of the filter. The maximal allowable time thus depends on the elevation angle and rate of change of elevation of the satellite and, just slightly, on the tide motion. In the case of interest, it turns out the maximum filtering time should be around 10 seconds. In other words, in 10 seconds, at least for one satellite, the coherent interferometric phase changes by  more than  $\frac{\pi}{2}$ radians. With this filter length it is not possible to separate coherent and incoherent components of the field, and  fading events are not completely eliminated. However,  a realistic estimation of the bridge height (and bias) does become possible. 

As mentioned, to extract the coherent component a longer averaging period should be used. To this end, we first  counter-rotate the interferometric field using a first guess of the bridge height, as we now explain in more detail.

After downconversion and despreading, we can express the reflected complex field as a sum of the coherent and incoherent components, 
\begin{equation}
E(t) = A_r\, e^{i{\cal L}(t)/\lambda }+O(t). 
\end{equation}
Now consider that we have a first guess for the height and bias parameters, i.e., a model for the lapse ${\cal L}_m$. This model is used to counter-rotate the field:
\begin{eqnarray}
 E^{cc}(t)&=& E(t)/E_m(t)  \\
%&= &\exp\{2\,h_b\left(t\right)\,\sin\left(\epsilon_{p}\left(t\right)\right)+{b}\}\cdot \nonumber \\
%&= & e^{-2i\,\hat{h}\left(t\right)\,\sin\left(\epsilon_{p}\left(t\right)\right)+\hat{d}_{offset}}
%\end{eqnarray}
%obtaining
%\begin{eqnarray}
&=& A_r \, e^ {2\,i\, \delta h(t) \,\sin\left(\epsilon_{p}\left(t\right)\right)+ %\nonumber \\
i\delta {b} }+O'(t). \nonumber
\label{eq:interf_field_crot}
\end{eqnarray}
Clearly, the phase of the coherent field in equation \eqref{eq:interf_field_crot} will now vary much slower than the phase of the original reflected field as a function of the elevation (and therefore time). This allows for a longer filtering time, and the exraction of the coherent component of the signal (recall that $O(t)$ has zero mean).

The equation which relates the counterotated phase lapse between direct and reflected signal, the satellite elevations and the $\delta h$ (i.e. the error between the first guess of the bridge height and the real value) is
\begin{equation}
{\cal L}_P^c(t) = 2\,\delta h  \left(t\right)\cdot\sin\left(\epsilon_{P}\left(t\right)\right) +\delta N_{P}\lambda+\delta {b}.
\label{eq:delayVSDeltah_phase}
\end{equation}
This is the new equation to be used to fit the lapse versus sine of elevation straight line and infer the height offset  of the bridge and bias (with respect to the first guess used to counter-rotate the field).

%As already said, the equation relating the interferometric phase and the bridge height is known up to a multiple of wavelength (equation \eqref{eq:delayVSh_phase}) or, in other words, there is an ambiguity due to the periodic behaviour of the phase. \
In order to solve the ambiguity problem, a search is performed in the space of the integer n-tuples  and the one that produces the linear fit with smallest residue is selected. It is important to point out that the  n-tuple search space is drastically reduced by the prior field counter-rotation. 
%In fact, since the only residual \it movement \rm in the filtered data is now the tide, the ambiguity search space   \it filtered \rm phase histories will all lie in a narrow height interval. This interval is determined basically by the correctness of first guess for the bridge height. 
For example, if the guess is within $\pm$ half meter, the n-tuple subspace to be scanned can be limited to those n-tuples whose components belong to the interval $[-3,3]$, centered on the first guess of the n-tuple, as obtained from a real (as opposed to integer) ambiguity resolution.

Another way to reduce the cardinality of the subspace of the n-tuples to check is to consider that satellites with similar elevation angles cannot have very different integer ambiguities.

\section{Results and comparison}
\label{ch:first_altimetric_results}
The PARFAIT  algorithm described in the previous section, has been used to analyze  the first 10 minutes of the Bridge-2 data, Part A1 and to the first 10 minutes of  Part A2. The following steps have been performed accordingly in batches of 2 minutes:
\begin{itemize}
\item The EM fields, direct and reflected, have been computed through the usual  correlation process.
\item The   complex interferometric field has been counter-rotated (Equation \eqref{eq:interf_field_crot}).
\item The counter-rotated  field has been filtered using a 30 s window.
\item The phase of the interferometric, counter-rotated and filtered field has been unwrapped.
\item For every possible set of values of the ambiguities  $N_{P}$, a straight line has been interpolated to the phase histories (one for each visible satellite) against the elevation angle (Equation \eqref{eq:delayVSDeltah_phase}). The best fit has been identified.
%\item The  n-tuple of values of $\lambda N_{p}$ minimizing the residue of the fit is use to define a first guess for the integer n-tuple of values of $N_{p}$;
%\item the linear fit has been again performed for every integer n-tuple of $N_{p}$ to find the one that minimises the residuals.
\end{itemize}

\begin{figure}[tbhp]
%\hspace{.5cm}
\centering
 \includegraphics[width=10cm]{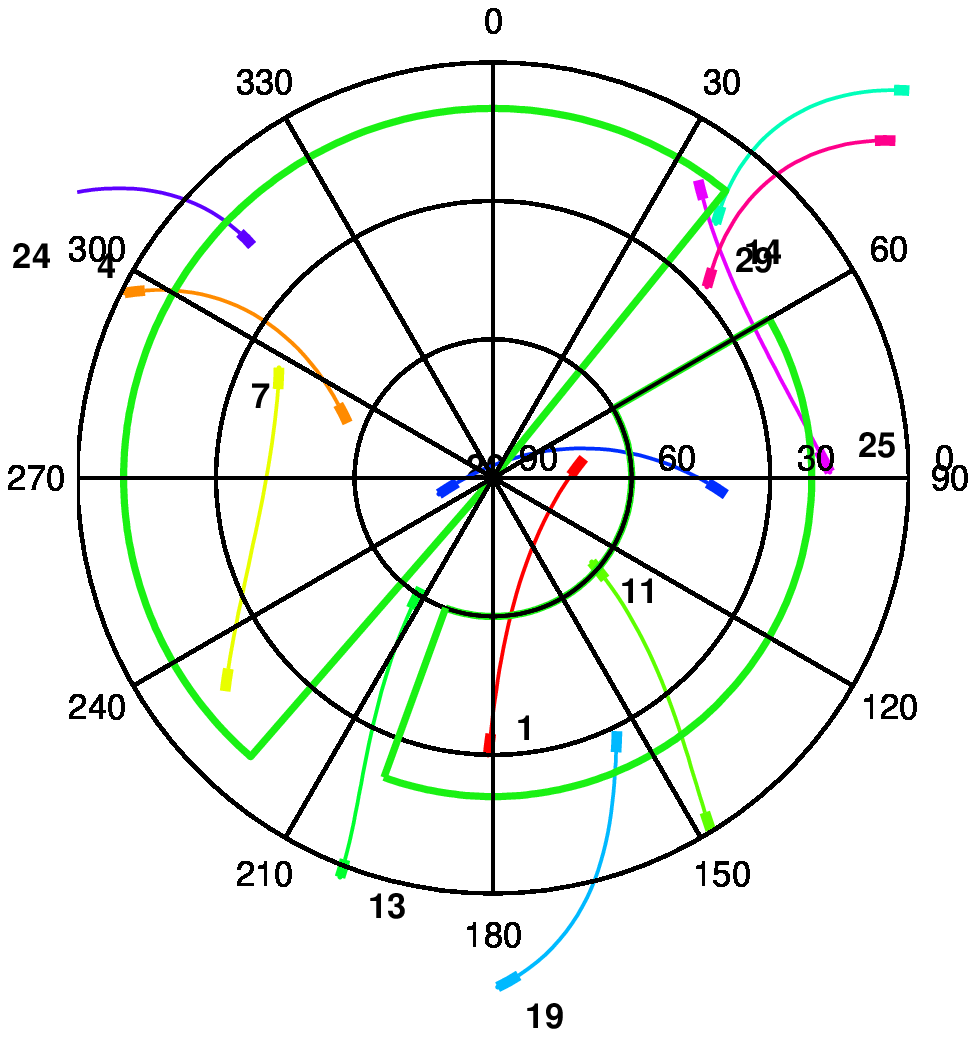}
 \caption{Each colored arc represents the position of a GPS satellite from the start of the Part A1 of the experiment to the beginning of Part A2 plus 10 minutes.  The green mask represent the area where the GPS signal reflections are supposed to be free of shadowing phenomena due to the bridge structure, and therefore only the satellite within this mask can be taken into consideration for PARIS processing. The bold parts of the lines represent the first and the second 10 minutes periods.}
\label{fig:zeeland_sat}
\end{figure}

\begin{figure}
  \centering
    \subfigure[In this figure, the reflected-minus-direct phase delay for each PRN is  plotted with $N_{P}=0$.]
    {\includegraphics[width=8cm]{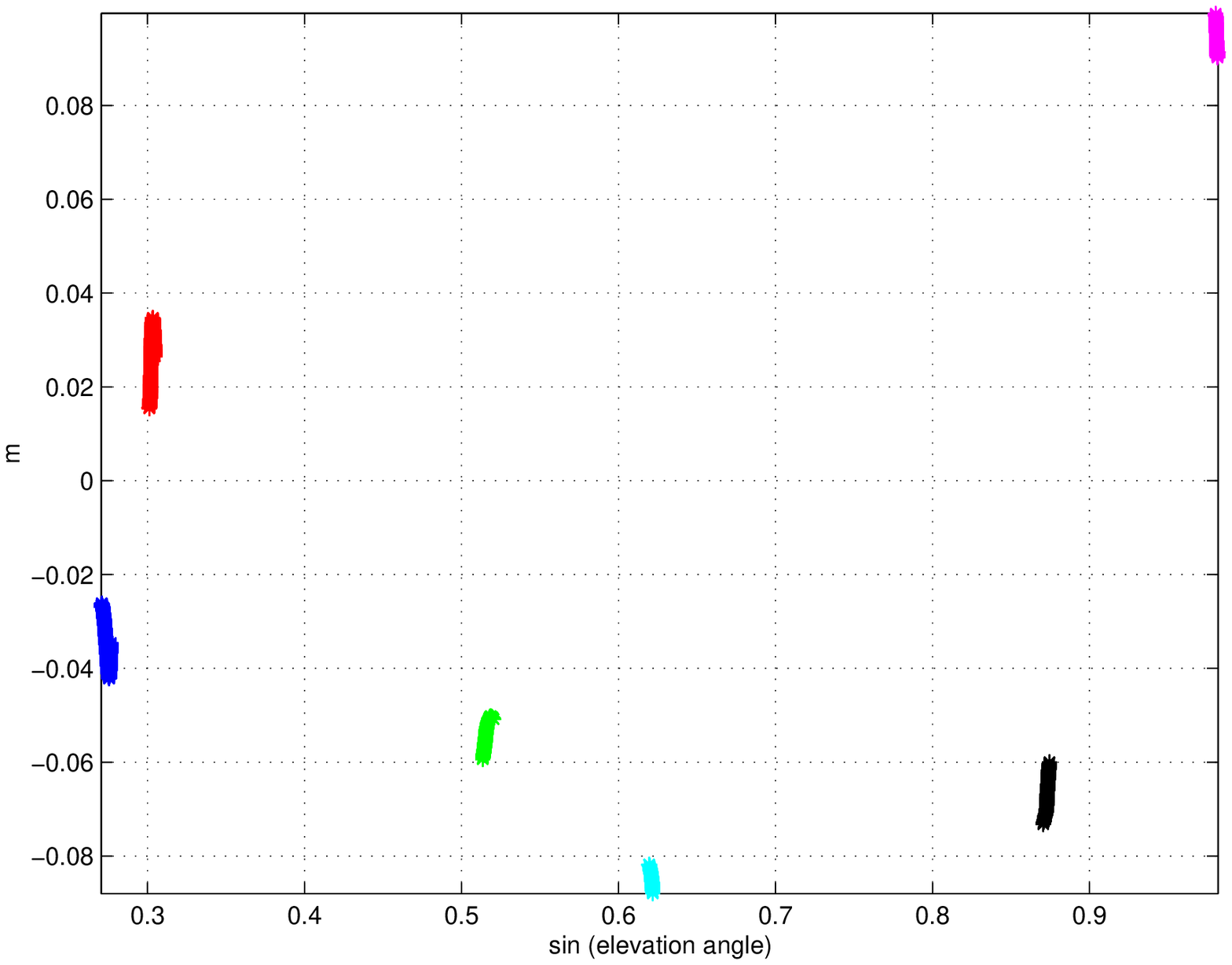}
      \label{fig:phases_hist_example_1}}
    %\hspace{1cm}
    \subfigure[In this figure, the reflected-minus-direct phase delay for each PRN is  plotted with $N_{P}=\{0 0 1 1 2 3\}$.]
    {\includegraphics[width=8cm]{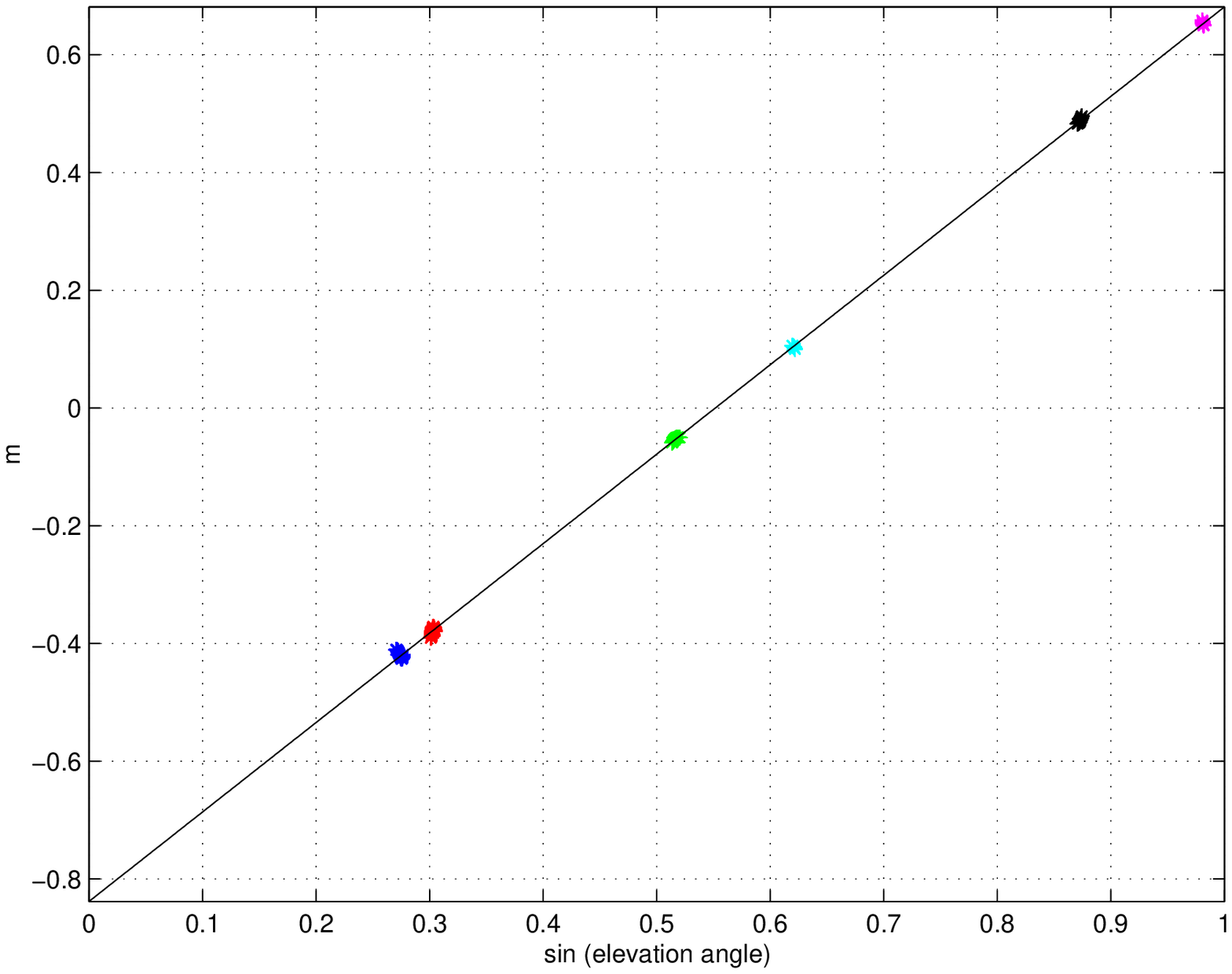}
      \label{fig:phases_hist_example_2}}
  \caption{Each colored spot represents the reflected-minus-direct phase delay versus satellite-elevation for a different satellite (PRN number).}
  \label{fig:SNR_examples}
\end{figure}

%The filter used for smoothing the field is a flat zero-phase filter, with a length of 30 seconds.\\
The analysis has been carried out for almost\footnote{Satellites outside the \it Zeeland Mask \em \cite{belmonte2002,belmonte2003} are not considered (see also the caption in Figure \ref{fig:zeeland_sat}).} all visible satellites (see Table 2 and Figure \ref{fig:zeeland_sat}). The phase histories are shown in Figure \ref{fig:phases_hist_example_1}. 
A straight line has been fit through these phase histories, against the sine of the satellite elevation angle (Figure \ref{fig:phases_hist_example_2}).

This fitted line gives an estimation of the bridge height of 18.61 m, a hardware bias of -0.81 m and, as first guess for the n-tuple that solves the ambiguity problem, the values $[0\:0\: 1 \:1\: 2\: 3]$. Now, a search in a subset of $\mathcal{I}^6$ is carried out to minimise the residuals of the fit in the space of the n-tuples of integers. The subspace considered is the one spanned by all the combination of integers between $\pm3$ around the first guess. The result is the n-tuple $[0\: 0\: 2 \:2 \:4\: 5]$ which gives a bridge height estimation of $18.82$ m and an instrumental bias of $-0.45$ m. 

\begin{table*}
\vspace{.5cm}
\label{tab:visible_sat1}
\centering{
\begin{tabular}{|r|c|c|c|}
  \hline
 PRN & elevation & \parbox{2cm}{\centering{mean $SNR$ (direct)}} & \parbox{2cm}{\centering{mean $SNR$ (reflected)}} \\
        \hline \hline
14  & 17$^o$   & 29.4 dB& 25.0 dB\\
  \hline                  
25  & 17$^o$   & 32.0 dB& 25.8 dB \\
  \hline                  
1  &  30$^o$   & 31.2 dB& 24.6 dB\\
  \hline                  
7  &  38$^o$   & 33.2 dB& 29.4 dB\\
  \hline                  
11  & 62$^o$   & 34.0 dB& 29.4 dB\\
  \hline                  
20  & 78$^o$   & 30.4 dB& 26.6 dB\\
      \hline
\end{tabular}}
\caption{Visible satellites, their elevation in degrees, the 10 ms coherent integration mean $SNR_{dB_w}$ ($20\log_{10}$[peak-grass/grass correlation coefficient]) for the direct and the reflected signal.}
\end{table*}

\begin{figure}
  \centering
    \subfigure[The upper line is the bridge height over the water line estimated using PARFAIT. The bottom line is the bridge altitude according to the available tide measurements and the calibration of the absolute height  using a GPS buoy \cite{belmonte2002}.]
    {\includegraphics[width=8cm]{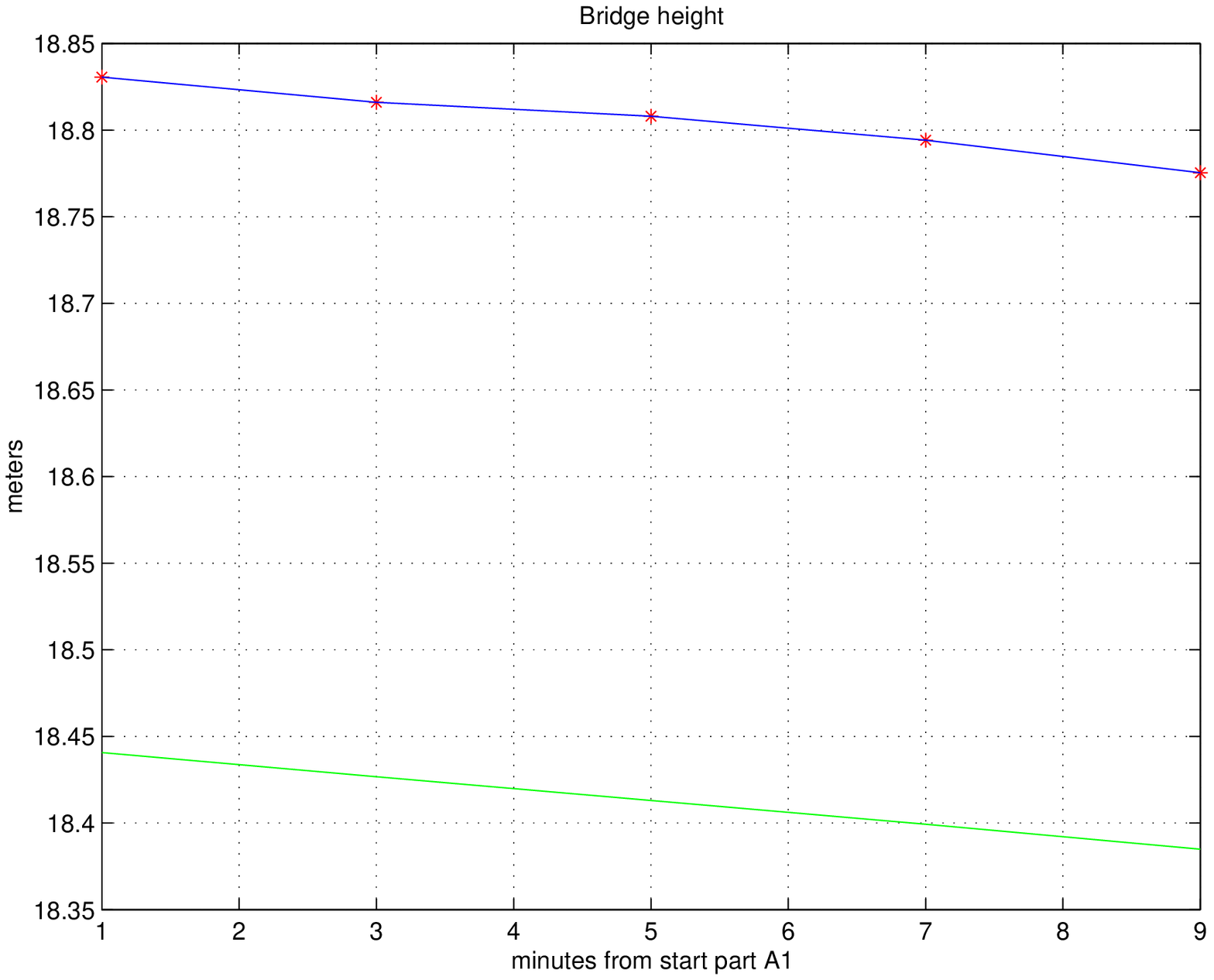}
      \label{fig:bridge_cfr_partA1_1}}
    %\hspace{1cm}
    \subfigure[In this plot, the estimates of the height (red *) and the actual height (line) are shown, after adding to the actual height the mean of the values of the last column of Table 3.]
    {\includegraphics[width=8cm]{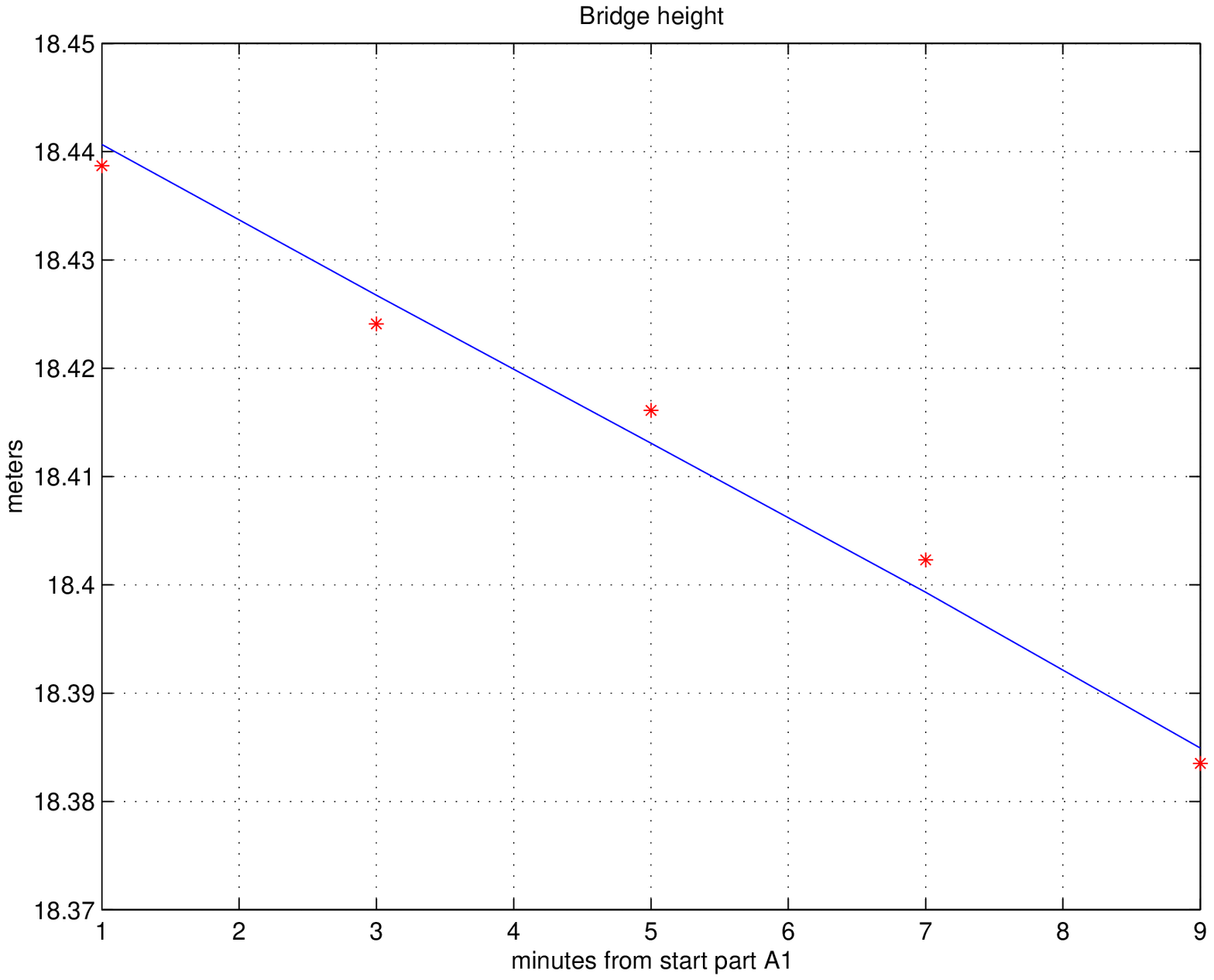}
      \label{fig:bridge_cfr_partA1_2}}
  \caption{The bridge height estimate during the first 10 minutes of the Part A1 of the data.}
  \label{fig:bridge_cfr_partA1}
\end{figure}

\begin{figure}
  \centering
    \subfigure[The upper line is the bridge height estimated using PARFAIT. The bottom line is the bridge height according to the available tide measurements and the GPS buoy measurement \cite{belmonte2002}.]
    {\includegraphics[width=8cm]{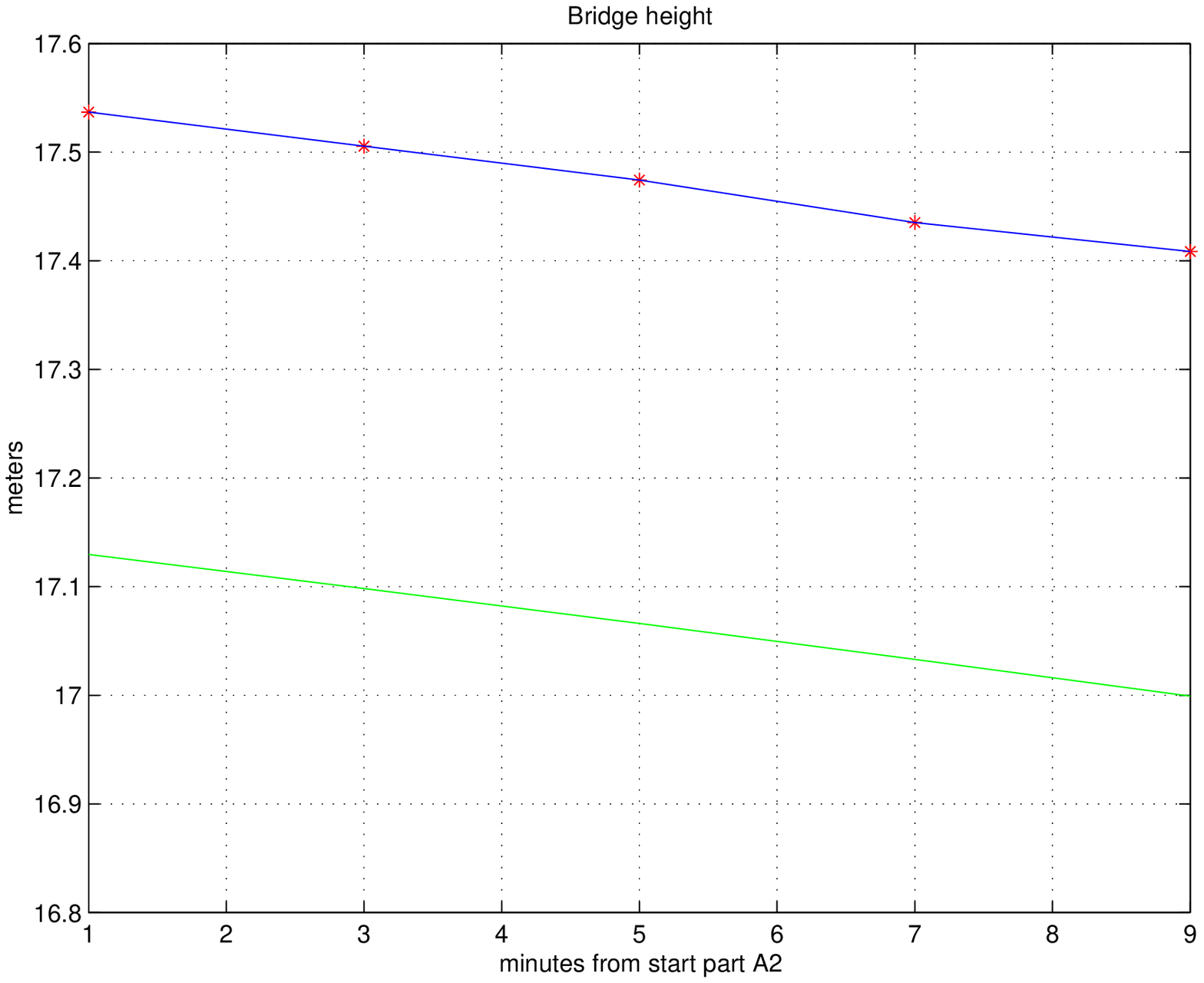}
      \label{fig:bridge_cfr_partA2_1}}
    %\hspace{1cm}
    \subfigure[In this plot, the estimates of the height (red *) and the actual height (line) are shown, after adding to the actual height the mean of the values of the last column of Table 4.]
    {\includegraphics[width=8cm]{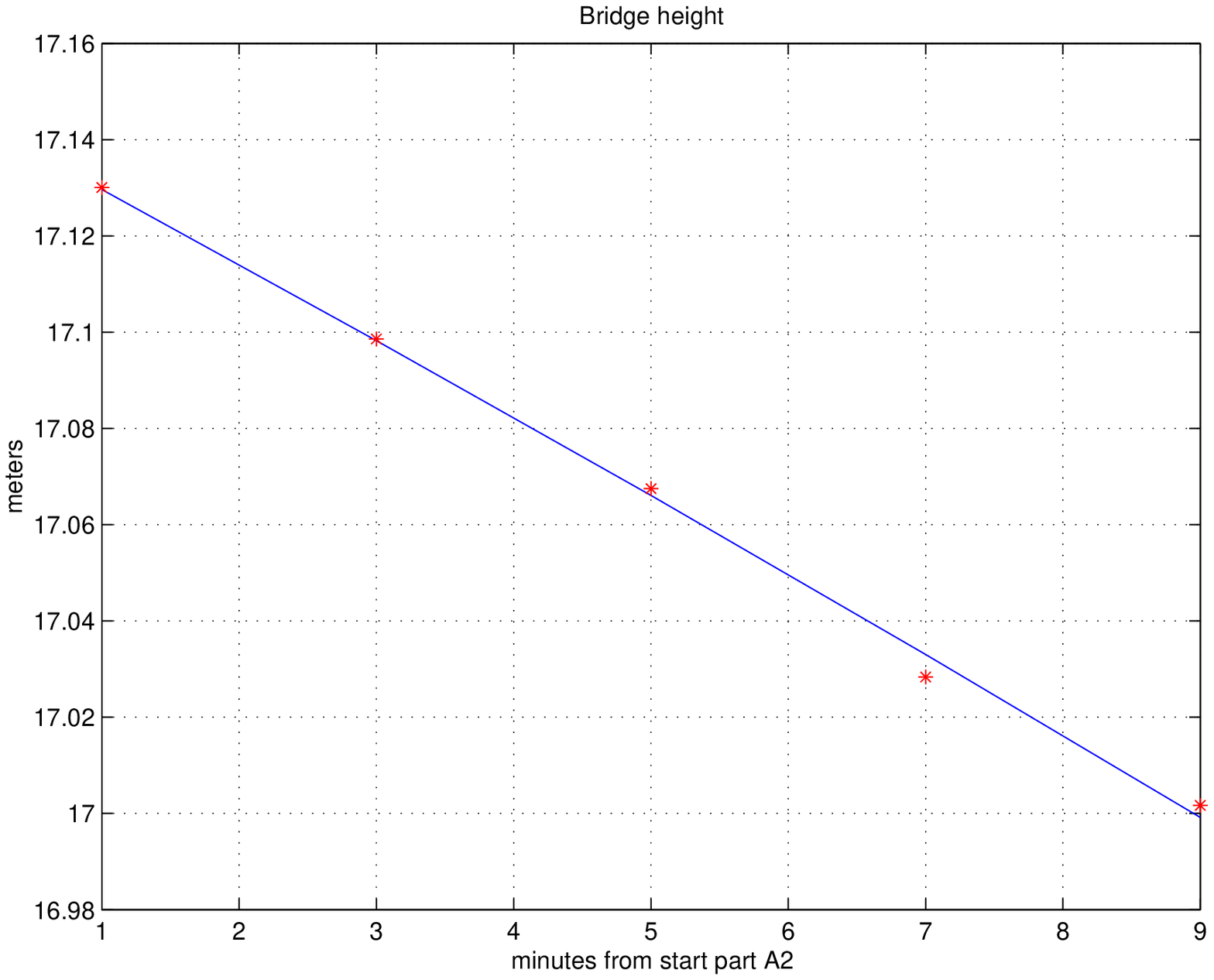}
      \label{fig:bridge_cfr_partA2_2}}
  \caption{The bridge height estimation during the first 10 minutes of the Part A2 of the data.}
  \label{fig:bridge_cfr_partA2}
\end{figure}

This procedure has been implemented with data from  the first 10 minutes of Part A1 and  A2 of the Bridge-2 experiment. The results are reported in Table 3 and in Figure \ref{fig:bridge_cfr_partA1} for Part A1 and in Table 4 and in Figure \ref{fig:bridge_cfr_partA2} for Part A2. 
%The final value of the estimated tide is considered to be the interpolated straight line through the available point \it after removing the supposed estimation bias \rm . 
%The use of a straight line for the interpolation is justified by the short period of time considered, relatively to the tide period. The standard deviation of the interpolated estimation w.r.t. the measured tide is of $0.30$ cm for part A1 and of $0.28$ cm for part A2. The results for both periods are plotted in figure \eqref{fig:true_bh_vs_est_bh}.\\

Fitting both parts to the tide curve, i.e. choosing the bias that minimizes the standard deviation of the data to the available tide ``ground truth'', leads to an altimetry bias of 40~cm and a standard deviation of less than 1~cm.
%0.83
This bias could have an origin in the ground ``truth'',  due either to an error in the determination of the absolute value of the height of the bridge performed using  the GPS buoy available data (only a few seconds, which may  have caused ambiguity resolution problems)  or, partially, to some anomalies in the water mass flow in the vicinity of the bridge structures.

Moreover, considering also that the tide dynamics measured below the bridge may be delayed with respect to the place were the tide was  measured, the best fit (over both bias and delay) is obtained with a delay of 1 minutes and 37 seconds with respect to the time of the tide data collection and with a bias of 40 cm. The standard deviation of the fitted data with respect to the tide curve is in this case of 0.3 cm. %0.28

\begin{table*}[ht]
\label{tab:height_results_1}
\centering{
\begin{tabular}{|c||c||c|c||c|}
  \hline
 \parbox{2cm}{\centering{time (minutes from start)}} & \parbox{2cm}{\centering{instrumental bias [cm]}} &\parbox{2.5cm}{\centering{bridge height estimation [m]}} &\parbox{2cm}{\centering{assumed ground truth [m]}} & \parbox{2cm}{\centering{difference [cm]}} \\
\hline \hline
1      & -0.45 &  18.83   &  18.44  & 39.71 \\
\hline                  
3      & -0.45 &  18.82   &  18.42  & 39.63 \\
\hline                  
5      & -0.46 &  18.81   &  18.41  & 40.19 \\
\hline                  
7      & -0.45 &  18.79   &  18.39  & 40.21 \\
\hline                  
9      & -0.26 &  18.78   &  18.38  & 39.83 \\
\hline 
\end{tabular}}
\vspace{.5cm}
\caption{Results of the bridge height estimation during the first 10 minutes of the Part A1 data. }
\end{table*}

%Another interesting check to be done to asses the validity of the processing is to compare the change in the height of the bridge between the first 10 minutes of part A1 and the first 10 minutes of part A2 as retrieved by the processing and as provided by the tide measurements. This comparison is shown in figure \eqref{fig:crf_with_tide_2}. The result is absolutely satisfying: the estimation is in accordance with the measures within about 4 cm.\\

 To summarize, the proposed approach to PARIS altimetry, the \sc parfait \rm  technique, leads a  very precise estimation of the tide, 
\begin{itemize}
\item without the need to insert any kind of model for the phase of the reflected signal during fadings, 
\item without rejecting too many visible satellites because of their poor SNR and/or frequent fadings.
%\item without the need to correct for the navigation message (being enough to be aligned with it).
\end{itemize}

Finally, we note that this technique is directly applicable for PARIS phase processing from air and spaceborne applications, as long as a suitable model for the lapse phase can be constructed. This will be the subject of future work.

\begin{table*}[ht]
\label{tab:height_results_2}
\centering{
\begin{tabular}{|c||c||c|c||c|}
  \hline
 \parbox{2cm}{\centering{time (minutes from start)}} & \parbox{2cm}{\centering{instrumental bias [cm]}} &\parbox{2.5cm}{\centering{bridge height estimation [m]}} &\parbox{2cm}{\centering{assumed ground truth [m]}} & \parbox{2cm}{\centering{difference [cm]}} \\
\hline \hline
1      & -0.27 & 17.54    & 17.11   & 42.34 \\
\hline         
3        & -0.28 & 17.52  & 17.08   & 42.36 \\
\hline         
5      & -0.26 & 17.47    & 17.04   & 42.52 \\
\hline         
7        & -0.08 & 17.44  & 17.02   & 41.95 \\
\hline         
9       & -0.08 & 17.41   & 16.98   & 42.71 \\
\hline 
\end{tabular}}
\vspace{.5cm}
\caption{Results of the bridge height estimation during the first 10 minutes of the Part A2 data. }
\end{table*}

%\pagebreak
%\clearpage

\section*{Acknowledgements} 

The authors wish to thank Manuel Martin-Neira (Technical Manager of the ESA/ESTEC Contract No. 14285/85/nl/pb under which this work was carried out) and Maria Belmonte (ESA/ESTEC) for useful discussions and real collaboration. We also thank the other partners in the project, especially GMV for the GPS buoy data analysis. 

{\em 
All Starlab authors have contributed significantly; the Starlab author list has been ordered randomly.}

%\pagebreak
%\clearpage
%\addcontentsline{toc}{chapter}{List of Figures }
%\listoffigures
%\pagebreak
%\addcontentsline{toc}{chapter}{List of Tables} 
%\listoftables

%\pagebreak
%\bibliographystyle{alpha}
\bibliographystyle{plain}
\bibliography{/home/alkaid/intranet/library/feosbiblio}
\addcontentsline{toc}{chapter}{Bibliography}

\end{multicols}

\ifpdf
\begin{figure}[tbhp]
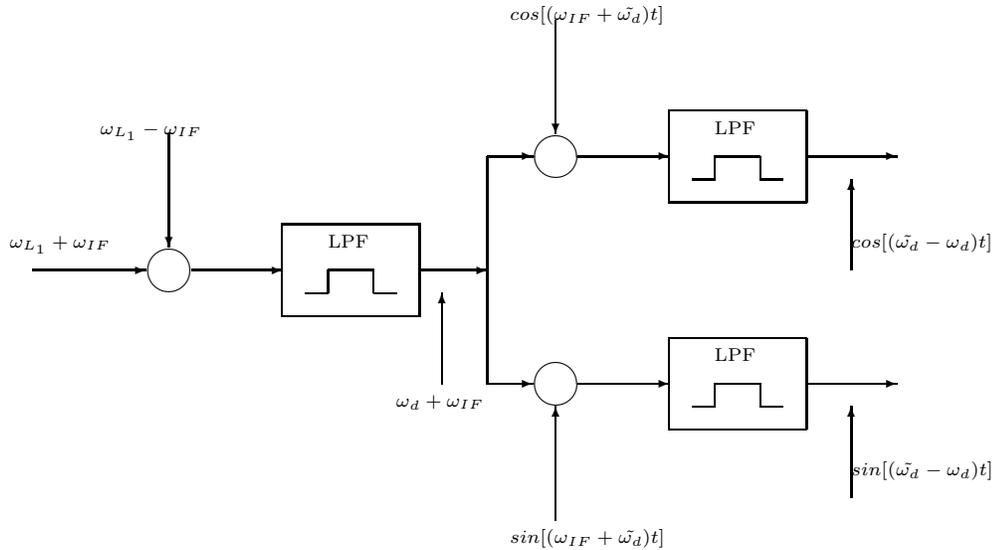

\centering
      \resizebox{10cm}{!}{
        \includePSTeX{downconvertion} 
}
    \caption{Starting from the in-phase component of the sampled signal, with a carrier frequency equal to the sum of the IF of the receiver and the Doppler frequency, a complex downconversion is performed. After  low-pass filtering, the two obtained signals bring information about the amplitude of the backscattered EM field for both the in-phase and in-quadrature components---the  spread, complex field.}
  \label{fig:downconvertion}
\end{figure}
\else
\begin{figure}[tbhp]
\centering
\setlength{\unitlength}{4144sp}%
\begingroup\makeatletter\ifx\SetFigFont\undefined%
\gdef\SetFigFont#1#2#3#4#5{%
  \reset@font\fontsize{#1}{#2pt}%
  \fontfamily{#3}\fontseries{#4}\fontshape{#5}%
  \selectfont}%
\fi\endgroup%
\begin{picture}(5320,3251)(1,-2496)
\thinlines
{\put(881,-766){\line( 1,-1){136.500}}
}%
{\put(881,-903){\line( 1, 1){136.500}}
}%
{\put(3195,-86){\line( 1,-1){136}}
}%
{\put(3195,-222){\line( 1, 1){136}}
}%
{\put(3195,-1575){\line( 1, 1){136.500}}
}%
{\put(3195,-1438){\line( 1,-1){136.500}}
}%
{\put(954,-830){\circle{272}}
}%
{\put(3268,-149){\circle{272}}
}%
{\put(3268,-1511){\circle{272}}
}%
{\put(137,-830){\vector( 1, 0){680}}
}%
{\put(954,-14){\vector( 0,-1){680}}
}%
{\put(2859,-149){\line( 0,-1){1362}}
}%
{\put(2859,-149){\vector( 1, 0){272}}
}%
{\put(2859,-1511){\vector( 1, 0){272}}
}%
{\put(3403,-149){\vector( 1, 0){545}}
}%
{\put(3268,667){\vector( 0,-1){681}}
}%
{\put(3403,-1511){\vector( 1, 0){545}}
}%
{\put(3268,-2327){\vector( 0, 1){680}}
}%
{\put(5037,-830){\vector( 0, 1){544}}
}%
{\put(3948,-1783){\framebox(817,545){}}
}%
{\put(4084,-1647){\line( 1, 0){136}}
\put(4220,-1647){\line( 0, 1){136}}
\put(4220,-1511){\line( 1, 0){272}}
\put(4492,-1511){\line( 0,-1){136}}
\put(4492,-1647){\line( 1, 0){136}}
}%
{\put(4765,-1511){\vector( 1, 0){544}}
}%
{\put(5037,-2192){\vector( 0, 1){545}}
}%
{\put(1634,-1103){\framebox(817,545){}}
}%
{\put(2451,-830){\vector( 1, 0){408}}
}%
{\put(1090,-830){\vector( 1, 0){544}}
}%
{\put(1771,-966){\line( 1, 0){135}}
\put(1906,-966){\line( 0, 1){136}}
\put(1906,-830){\line( 1, 0){273}}
\put(2179,-830){\line( 0,-1){136}}
\put(2179,-966){\line( 1, 0){136}}
}%
{\put(2587,-1511){\vector( 0, 1){545}}
}%
{\put(3948,-422){\framebox(817,545){}}
}%
{\put(4084,-286){\line( 1, 0){136}}
\put(4220,-286){\line( 0, 1){137}}
\put(4220,-149){\line( 1, 0){272}}
\put(4492,-149){\line( 0,-1){137}}
\put(4492,-286){\line( 1, 0){136}}
}%
{\put(4765,-149){\vector( 1, 0){544}}
}%
\put(545,-14){\makebox(0,0)[lb]{\smash{\SetFigFont{7}{8.4}{\rmdefault}{\mddefault}{\updefault}$\omega_{L_1}-\omega_{IF}$}}}
\put(4220,-1375){\makebox(0,0)[lb]{\smash{\SetFigFont{7}{8.4}{\rmdefault}{\mddefault}{\updefault}LPF}}}
\put(5037,-694){\makebox(0,0)[lb]{\smash{\SetFigFont{7}{8.4}{\rmdefault}{\mddefault}{\updefault}$cos[(\tilde{\omega_d}-\omega_d)t]$}}}
\put(1906,-694){\makebox(0,0)[lb]{\smash{\SetFigFont{7}{8.4}{\rmdefault}{\mddefault}{\updefault}LPF}}}
\put(2995,-2464){\makebox(0,0)[lb]{\smash{\SetFigFont{7}{8.4}{\rmdefault}{\mddefault}{\updefault}$sin[(\omega_{IF}+\tilde{\omega_d})t]$}}}
\put(5037,-2056){\makebox(0,0)[lb]{\smash{\SetFigFont{7}{8.4}{\rmdefault}{\mddefault}{\updefault}$sin[(\tilde{\omega_d}-\omega_d)t]$}}}
\put(  1,-694){\makebox(0,0)[lb]{\smash{\SetFigFont{7}{8.4}{\rmdefault}{\mddefault}{\updefault}$\omega_{L_1}+\omega_{IF}$}}}
\put(2995,667){\makebox(0,0)[lb]{\smash{\SetFigFont{7}{8.4}{\rmdefault}{\mddefault}{\updefault}$cos[(\omega_{IF}+\tilde{\omega_d})t]$}}}
\put(4220,-14){\makebox(0,0)[lb]{\smash{\SetFigFont{7}{8.4}{\rmdefault}{\mddefault}{\updefault}LPF}}}
\put(2315,-1647){\makebox(0,0)[lb]{\smash{\SetFigFont{7}{8.4}{\rmdefault}{\mddefault}{\updefault}$\omega_d+\omega_{IF}$}}}
\end{picture}
    \caption{Starting from the in-phase component of the sampled signal, with a carrier frequency equal to the sum of the IF of the receiver and the Doppler frequency, a complex downconversion is performed. After  low-pass filtering, the two obtained signals bring information about the amplitude of the backscattered EM field for both the in-phase and in-quadrature components---the  spread, complex field.}
  \label{fig:downconvertion}
\end{figure}
\fi


\begin{figure}[t!]
%\hspace{.5cm}
\centering
 \includegraphics[width=8cm]{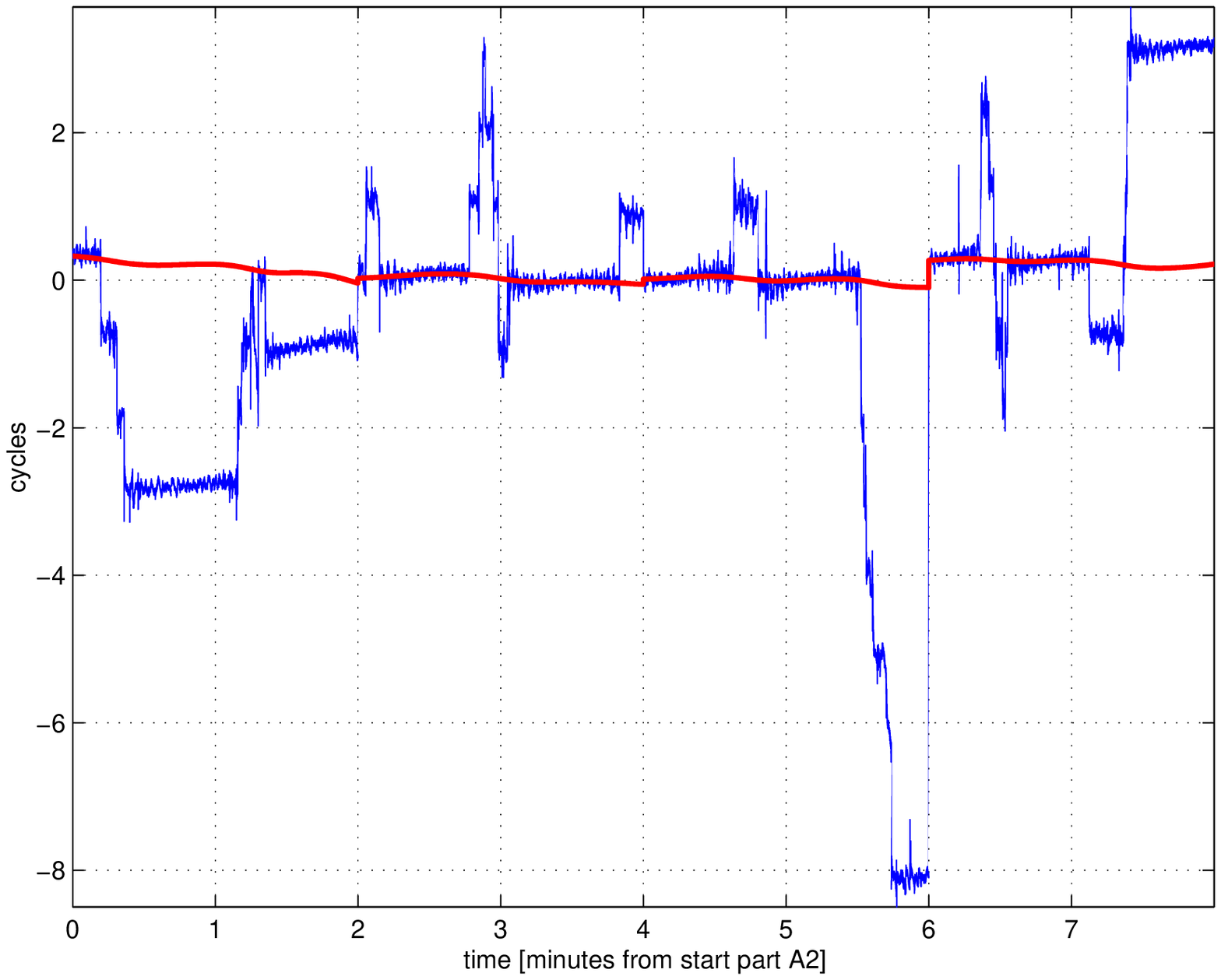}
 \caption{ \label{fig:filtered_phase} The blue curve is the phase of the interferometric field for PRN number 7, from minute 0 to minute 8 of  Part A2 of the Bridge 2 experiment. The occurrence of a fadings and  of isolated cycle slips can be seen. These phenomena disappear in the phase of the filtered interferometric field (red line).}
\end{figure}

%\begin{figure}[t!]
% \centering
%\label{fig:crf_with_tide_2}
% %\hspace{0.5cm}
% \includegraphics[width=8cm]{crf_with_tide_2}
% \caption{This line represents the difference of two differences: the difference of the measured bridge heights during the collection of the first part of data and those measured during the second part minus the same difference but calculated with the estimated data. The result confirms again the effectiveness of the \sc parfait \rm  processing.}
%\end{figure}

\begin{figure}[t!]
 \centering
\label{fig:true_bh_vs_est_bh}
 %\hspace{0.5cm}
 \includegraphics[width=14cm]{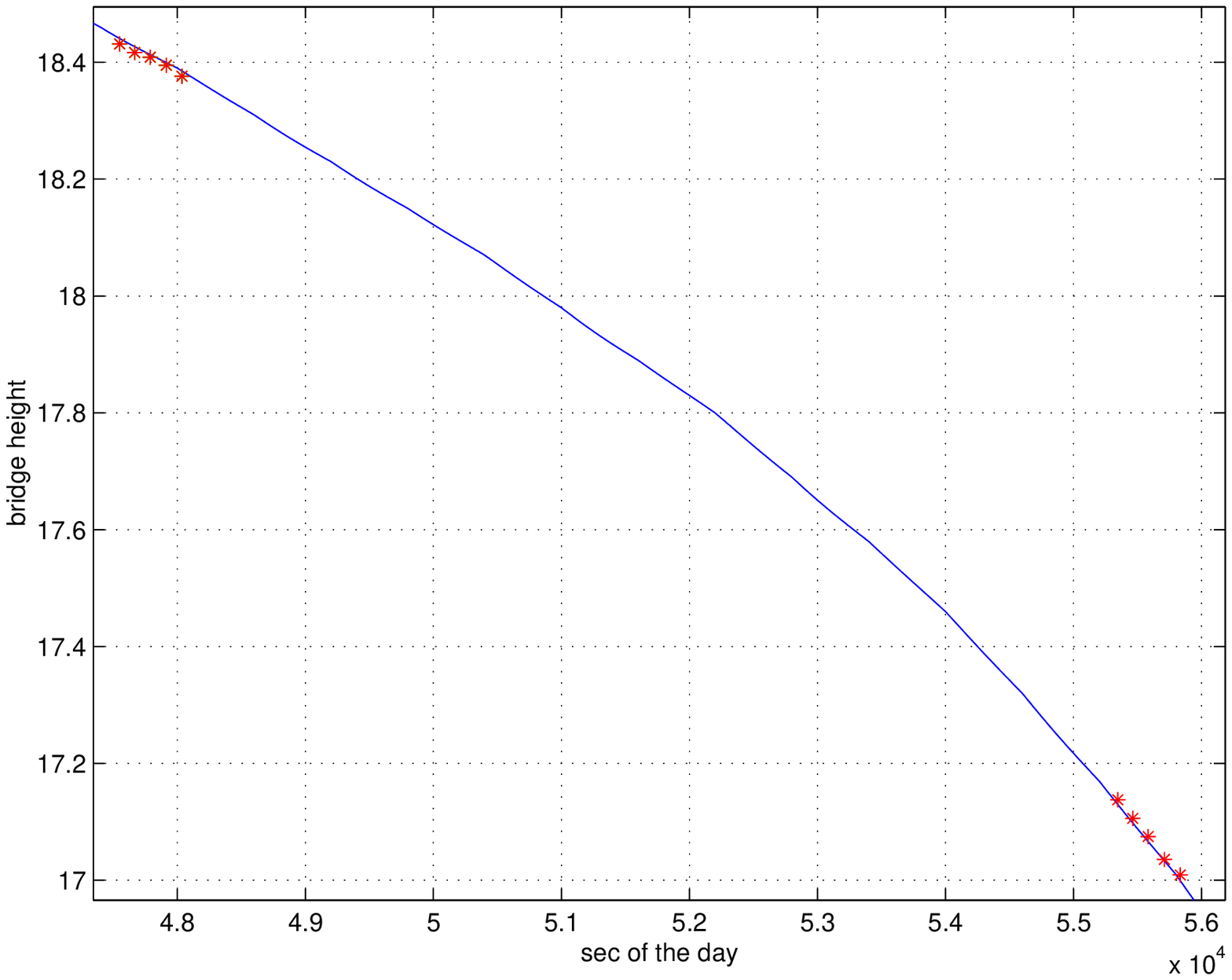}
 \caption{ The solid line is the distance between the up-looking antenna and the sea surface, according to the available tide measurements and the GPS buoy measurement \cite{belmonte2002}. The green dots are the estimated values, after removing the vertical bias.}
\end{figure}

\end{document}
\end